\documentclass[preprint,review,12pt,a4paper,fleqn]{cas-sc}

\usepackage{lineno}
\usepackage{amsmath}
\usepackage{caption}
\usepackage{booktabs}
\usepackage{amsfonts}
\usepackage{algorithm}
\usepackage[noend]{algpseudocode}
\usepackage{longtable}
\usepackage{multicol}
\usepackage{rotating}
\usepackage{placeins}
\usepackage{multirow}
\usepackage{subfig}
\usepackage{enumitem}
\usepackage{ulem}

\usepackage[authoryear]{natbib}
\usepackage{inconsolata}

\newcommand\red[1]{{\color{red}#1}} 
\newcommand\blue[1]{{\color{black}#1}} 
 
\usepackage{pgfplots} 
\pgfplotsset{compat=1.16} 

\def\tsc#1{\csdef{#1}{\textsc{\lowercase{#1}}\xspace}}
\tsc{WGM}
\tsc{QE}
\tsc{EP}
\tsc{PMS}
\tsc{BEC}
\tsc{DE}

\begin{document}

\let\WriteBookmarks\relax
\def\floatpagepagefraction{1}
\def\textpagefraction{.001}

\title[mode=title]{Waste Bins Location Problem: a review of recent advances in the storage stage of the Municipal Solid Waste reverse logistic chain}
\shorttitle{Waste bins location problem: a review}

\author[1,2]{Diego Gabriel Rossit}
[orcid=0000-0002-8531-445X]
\cormark[1]
\ead{diego.rossit@uns.edu.ar}

\address[1]{Department of Engineering, Universidad Nacional del Sur, Argentina}
\address[2]{Instituto de Matem\'{a}tica de Bah\'{i}a Blanca UNS-CONICET, Argentina}

\author[3]{Sergio Nesmachnow}[orcid=0000-0002-8146-4012]
\ead{sergion@fing.edu.uy}
\address[3]{Universidad de la República, Uruguay}

\cortext[1]{Corresponding author}
\shortauthors{D.~Rossit and S.~Nesmachnow}


\begin{abstract}
Municipal Solid Waste systems have important economic, social, and environmental impacts for society. Within the diverse stages of the Municipal Solid Waste reverse logistic chain, the waste bin location problem consists in properly locating bins in the corresponding urban area to store waste produced by the citizens. This stage has a large impact in the overall efficiency of the whole system. Thus, several researchers have addressed the location problem considering different optimization criteria and approaches. This article presents a comprehensive review of recent advances on the Waste Bins Location Problem, with the main goal of serving as a reference point for decision-makers in this area. 
\blue{The main findings indicate that several optimization criteria and resolution approachas have been applied, but few proposals have simultaneously optimized bins location and waste collection, or considered uncertainty of the model parameters and integrated approaches.}
\end{abstract}

\begin{graphicalabstract}
\includegraphics[width=1\textwidth,trim={2cm 0cm 2cm 0cm,clip}]{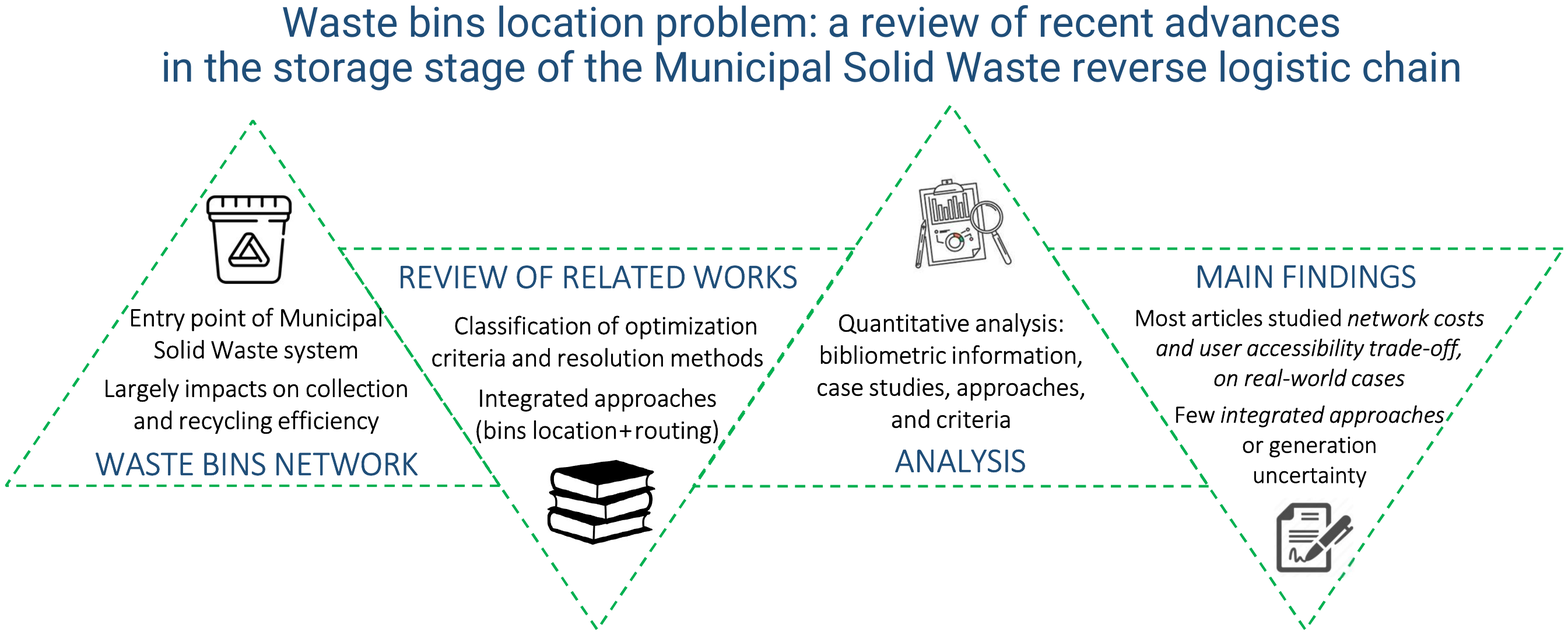}
\end{graphicalabstract}

\begin{highlights}
\item \blue{A comprehensive review of the waste bins location problem.}
\item Configuration of the network of bins affects the overall efficiency of the \blue{Municipal Solid Waste} system.
\item Waste bins location problem is a computationally complex problem.
\item Different conflicting criteria are involved in waste bins location.
\item \blue{Several approaches considered the trade-off between the cost of the system and quality of service}.
\item \blue{Few works addressed simultaneously the bins location and waste collection problems.}
\end{highlights}


\maketitle

\section{Introduction}
\label{sec:intro}


Waste management is a major challenge in modern societies, critical for sustainability and improving the quality of life of citizens~\citep{deX2017technologies}. This issue becomes even more critical if it is considered that the amount of waste generation rate per capita will continue increasing in the following decades~\citep{hoornweg2012waste,hoornweg2015peak}.

The reverse logistic chain of Municipal Solid Waste (MSW) is a special case of a reverse supply chain that is defined as ``a network consisting of all entities involved in the flow of disposed products leaving the point of consumption. It includes collection, transportation, recovery and disposal of waste. Its purpose is to recapture or create value and/or proper disposal''~\citep{van2020literature}. This subject has received a large interest from both professionals and academic communities in the last decades. An efficient provision of MSW management services is a key element to diminish the environmental impact of human activities on the environment, through a responsible a treatment and/or final disposition of the generated waste, and to postpone the depletion of limited resources, through the recovery of reusable resources from waste~\citep{das2019solid}.

The MSW reverse logistic chain includes several stages: generation, storage, collection, recovery, and/or disposal of waste. The storage stage consists in properly locating waste bins in the corresponding urban area to build a network for collecting the waste produced by the generators, which is the entry point to the MSW system. The location of waste bins has a large impact in the overall efficiency of the whole MSW system, since it influences the amount of waste that is introduced to the system, i.e., the material that circulates through the reverse logistic chain. A sparse or not properly distributed collection network reduces the accessibility of citizens to the MSW system, thus, reducing the amount of waste that is correctly deposed~\citep{parrotX2009municipal,toutouh2020soft}. Additionally, due to the direct influence of waste bins location and storage capacity on the length, travel times, and schedule of the collection routes, the design of this stage also contributes to improving the overall environmental impact of the MSW system~\citep{perez2017methodology,zhang2021sustainable}. The number, distribution, and type of waste bins are also relevant for the costs of the system since: the waste bins collection network and waste transportation from bins to transfer/final plants or landfills can explain about 70\% of the overall cost of the system~\citep{boskovic2016calculating}.

Waste bins location also has important social implications, since waste bins are usually the main interaction point between citizens and the MSW system of a city. For example, small problems in the management of this network--e.g., overflow of bins due to improper capacity planning--will immediately affect the quality of service provided to the citizens and may cause a large number of complaints. Additionally, waste bins, as other semi-obnoxious facilities, are affected by the ``Not In My Back Yard'' (NIMBY) phenomenon: citizens want bins to be located relatively near to minimize the distance they have to carry their waste to dispose it, but simultaneously they do not want them to be located very close to their homes to avoid unpleasant environmental consequences, e.g., bad smell, visual pollution or collection vehicle disturbing noises~\citep{coutinhoX2012bi}. Due to all the aforementioned economic, environmental and social aspects, waste bins location emerged as a relevant problem for designing efficient MSW systems, and several approaches have been proposed by the research community to solve it efficiently via different optimization approaches.

In this line of work, this article contributes with a thorough revision of the recent literature on the municipal solid waste bins location problem in urban areas. No similar survey has been published in the related literature. The only previous effort to review relevant works on the topic was elaborated by \citet{purkayastha2015collection}, focusing on the optimization of collection bin and recycle bin location. That review commented the main aspects and contributions of 17 articles (9 considering collection bins and 8 considering recycling bins) published between 1999 and 2014. Although it was a contribution to the area, being the first review on the topic, some results were not commented, and the main relevance of each proposal was not highlighted. Additionally, the review lacked of a global analysis of the research area, suggestions for interesting developments, or open research lines. In addition, many new researches about MSW bins location have been developed in the last six years, which are the main subject of the review presented in our article. We review more than seventy articles directly related to the MSW bin location problem in the period 1998--2021.

Thus, the main goal of the survey is to present an updated analysis of relevant articles about the MSW bin location problem to be considered as a guide for both researchers and practitioners involved MSW management and optimization. We put special emphasis on reviewing recently proposed approaches as well as considering new upcoming topics, such as the integration of the MSW bin location problem with the waste collection problem and the application of Internet of Things (IoT) technologies to waste bins.

The article is organized as follows. Next section describes the research methodology and 
\blue{content organization}.
\blue{Then, Sections~\ref{sec:reverse_supply_chain} and~\ref{sec:bin_location} present the main concepts of the overall reverse supply chain of MSW and of the specific stage of this chain regarding the waste bins location problem. Section~\ref{sec:quantitative} provides a quantitative summary of reviewed articles. The main concepts about bin location problem optimization criteria and the applied resolution methods are described in Sections~\ref{sec:objective_criteria} and~\ref{sec:resolution_method}.}
Finally, the discussion about relevant issues on the reviewed articles, open lines of research, and promising lines for future work are commented in Section~\ref{sec:conclusion}.

\section{Research methodology and content organization}
\label{sec:research_methodology}

Systematic literature review is a very relevant foundation for research, a must for knowing the main developments on any research area to properly contextualize the main contributions of the own research~\citep{Gough2012}. The main goal of a systematic literature review is to provide readers the current state-of-the-art regarding a topic or a research question, and how previous advances can be applied in a current of future situation. The existing knowledge must be properly described, synthesized, commented, and fairly evaluated using a sound and rigorous procedure~\citep{Snyder2019}. Classifications and conceptual categorizations are also useful to help the readers to better comprehend the main similarities and differences between the reviewed proposals~\citep{Tranfield2003}.

The study is within the category of \textit{systematic search and review}~\citep{Grant2009},
including 
two components: i) a comprehensive mapping, categorization, and quantitative/qualitative analysis of existing 
articles,
and ii) a critical review explaining the 
reported approaches and results.

The first stage in the literature review process is defining a proper research topic or research question, which can be properly used to guide the content and focus of each review. In the case of our study, the main research topic is the waste bins location problem, i.e., deciding the places in an urban area to deploy a set of bins in order to receive the waste from nearby generators.

The second stage is performing a systematic and comprehensive search to identify 
articles, reports, and thesis that most contribute to the area. The availability and reliability of data sources are crucial for a high-quality review. 
Our study considered the two main databases of scientific articles:
i) Scopus, 
(Elsevier Publishers, more than 36,000 peer-reviewed journals indexed)
and ii) Web of Science  
(Institute for Scientific Information/Clarivate Analytics, more than 35,000 journals indexed).
\blue{Since the search covered the main scientific publications in the subject, we have confidence that all main research approaches and ideas for waste bins location are represented, and the risk of bias is minimum.} 
Regarding the search criteria, the \texttt{\small TITLE-ABS-KEY(list of terms)} schema was 
applied, using as
keyword terms `waste', `bin', `location', `allocation',
and alternative words for bins (e.g., container, collection point).
No other (specific) terms were considered as keywords \blue{in order to reduce the risk of bias}. The search returned an initial set of 212 documents.

The third stage 
consists in filtering relevant articles, 
applying expert knowledge and pre-determined criteria for eligibility and relevance.
Each article was carefully examined, by reading its indexing information 
to determine its relevance. 
Many works were discarded, mainly  
unrelated, outdated articles, and those that report non-formalized, limited, or non-realistic studies. 
For example, articles focused on other stages of the MSW system, considering other types of waste (industrial, hospital, electronic waste), or describing waste monitoring and management systems were excluded since they are not directly related to the waste bins location problems.
A total number of 76 works were found to be directly related to the waste bins location problem.

The next stage involved properly organizing the selected related works into categories. According to the main goals of the survey and existing research initiatives, three main categories were identified, focused on specific details of the problem, methods, and problem variants addressed. Thus, after introducing a description of the reverse supply chain of MSW and the main details of the bin location problem, the review presents a summary and description of the optimization criteria considered for finding the most suitable locations for waste bins in urban areas. After that, a review of the different resolution approaches for solving the optimization problem reported in the bibliography are presented, aimed at identifying those most used and more promising resolution methods. Articles withing the category of \textit{integrated approaches} include proposals that solve the waste bins location problem simultaneously with other related optimization problems of the MSW system, mainly the design of waste collection routes. Articles within those categories are properly analyzed by quantitative indicators, and finally the main findings of the review are commented, aiming at identifying the main research trends, recent innovative approaches, and interesting open issues that may constitute future lines for research.

\section{The reverse supply chain of MSW}
\label{sec:reverse_supply_chain}

\begin{figure}[!h]
\setlength{\abovecaptionskip}{-20pt}
\setlength{\belowcaptionskip}{3pt}
\centering
\includegraphics[width=0.9\textwidth]{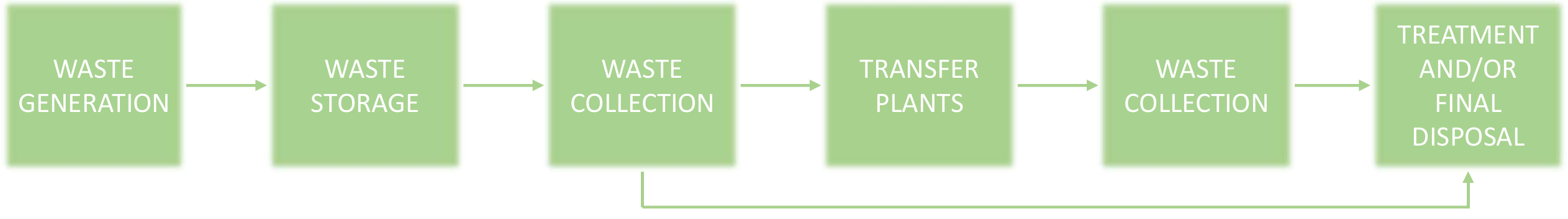}
\caption{Schema of the waste flow in the reverse logistic chain of MSW.}\label{fig:logistic_chain}
\end{figure}

Conversely to the concept of traditional (forward) logistics, reverse logistics implies, first, the physical transport of the used products from the final consumer to recovery/disposal points, and second, their transformation
into products that can rejoin the production system or, failing that, make an adequate final disposition \citep{fleischmannX1997quantitative}. According to this definition, 
the MSW system is a clear example of a reverse logistic chain~\citep{rossit2018thesis, van2020literature}. The reverse logistic chain of municipal waste involves several stages, as described in Figure~\ref{fig:logistic_chain}.

In the MSW logistic chain, waste is generated at households, commercial businesses, or institutions. Then, there is a storage stage that has its own characteristic depending on the collection system. The most common systems are:

\begin{itemize}[topsep=0pt,itemsep=2pt,partopsep=4pt,parsep=4pt]
    \item \textit{House-to-house or kerbside collection}. Collectors visit each generator in order to pick up the waste. The storage take place inside the generator and it is suppose to be performed on a short period of time since households do not usually have a large storage capacity~\citep{hoornweg2012waste}.
    \item \textit{Community bins}. Generators take their waste to some community bins, usually arranged in what is known as garbage accumulation points (GAPs)~\citep{tralhaoX2010multiobjective}. These GAPs are usually located in accessible places not very far from the generators. The garbage is stored there until it is picked up by collectors according to a certain schedule~\citep{hoornweg2012waste}.
    \item \textit{Self delivered system}. Generators carry their waste to disposal sites or transfer stations~\citep{hoornweg2012waste}. This is usually implemented for specific kinds of waste, such as hazardous waste, bulky waste and yard waste produced at households~\citep{dahlen2010evaluation} and the main difference with the community bins is that the facilities are usually larger and they provide some additional services than only been an storage point in the supply chain, e.g., a initial treatment process (compaction, separation) or even final disposition.
\end{itemize}

The decision among 
systems is site-specific.
The related literature has described
benefits for the community bins system over the kerbside collection, e.g., 
smaller cost in the collection phase
\citep{bonomoX2012method,gilardino2017combining} 
or are more suitable for collecting source classified waste~\citep{valeo1998location}. Moreover, the different storage systems are not mutually exclusive. For example, some cities use a kerbside collection for a certain type of waste, usually non-recyclable perishable waste that needs to be disposed relatively fast, an a community bins system for the waste that is not required to be treated immediately~\citep{dahlen2010evaluation}. 

The logistic chain of MSW continues with the collection of waste. Depending on the city and the characteristic of the MSW system, the logistic chain can include intermediate locations, called waste transfer stations. A transfer station is a facility, located near the generators, to receive and store waste from collection vehicles, before compacted to reduce volume and moved to larger vehicles for transportation to distant landfills or treatment centers~\citep{yadav2016feasibility}. Transfer stations can also include classification or recycling systems~\citep{eshet2007measuring}. A transfer station can be installed depending on the distance between generators and the landfill/treatment plant, the unit haul costs, and the economics of the facilities~\citep{chatzouridis2012methodology}, or the waste can be directly transported from generators to final disposition centers with the same vehicles (represented in Figure~\ref{fig:logistic_chain} by the bypass between the third stage waste collection and the sixth stage treatment and/or final disposal). Finally, depending on the type of waste and the level of development of the MSW system, the last stage in the MSW reverse logistic chain is treatment and/or final disposal. For treatment, incinerators, waste-to-energy plants, reclamation plants, or composite plants are used. In turn, disposal usually involves land filling or land spreading~\citep{ghianiX2014operations}.

\section{The waste bins location problem}
\label{sec:bin_location}

\blue{This section describes the waste bins location problem and relevant considerations.}

\subsection{\blue{Problem description}}
Among the several decisions involved in a MSW system, the design of the waste collection network is of capital importance since it is the entry point to the MSW system. A poor planning on this stage can diminish the amount of waste entering to the formal MSW system of a city. 

As the entry point to the system, the waste bins location problem 
is an example of the convergent design that is characteristic from reverse logistic networks, in which the material flow goes from many sources to few destination, in contrast to the divergent design of forward logistics, in which the material flow goes from a few sources to many destinations. In this case, the waste (material flow) is taken from many households (sources) to a few waste bins (destinations)~\citep{bingX2016research}. The waste bins location problem involves finding the best locations for community bins in an urban area while optimizing some relevant criteria, usually related to the cost of operating the system, the Quality of Service (QoS) provided to the citizens, or to both of them in a multiobjective fashion. 

Several stages of the MSW reverse logistic chain are affected by the decisions and strategies implemented in the waste bins location stage. The number of collection points, its distribution in the field, the type and size of the containers used, and the collection frequency are conditioning factors of the overall efficiency of the system~\citep{hazra2009solid,vijay2005estimation}. 
The location and configuration of collection points strongly influence the operational routing cost of waste collection from bins to the disposal sites~\citep{barrena2020solidarity,vu2018parameter}, and also impact on strategic levels of the reverse logistic chain, e.g., the designed capacity of intermediate and processing facilities, and the efficiency of the whole MSW system to collect all the generated waste~\citep{kumar2009characterization}. 
\citet{parrotX2009municipal} found that when the average distance between waste generators 
and bins
increases, the proportion of the population that uses containers decreases. This phenomenon causes an undesirable increase in waste accumulated in unsuitable places, such as waterways or green areas. In addition to the 
effects on the environment, low bin utilization also affects the population in more indirect ways through, 
e.g.,
dissemination of vector-borne diseases such as malaria or dengue~\citep{ali2019spatial,gupta2019use}, or increase in public charges for 
expenses to remove improperly disposed waste, to reincorporate it into the formal MSW system~\citep{parrotX2009municipal}. There is also evidence that a correct arrangement of bins in the urban area can encourage the community to correctly classify the waste at source, which usually is closely associated with the success of recycling programs~\citep{kao2013spatial,leeabai2019effects,leeabai2021effects}. As a relevant example, based on regression models developed from data collected in various cities in Spain, \citet{gallardo2010comparison} found that the decrease in the average distance between generator and waste disposal sites positively affects the percentage of differentiated waste that is collected. 

\subsection{\blue{Desirable characteristics of solutions}}

Besides its relevance for the MSW logistic chain, another aspect that have attract attention to the waste bins location problem is the difficulty to obtain good compromising solutions. Firstly, because in terms of computational complexity, it is an extension of the Capacitated Facility Location Problem, which was proven to be NP-hard~\citep{cornuejolsX1991comparison}, i.e., at least as hard as problems for which no efficient algorithms, which execute in polynomial time with respect to the size of the problem, have been devised. This class characterizes those problems that are difficult to solve computationally. Secondly, because locating bins is hard because of the conflicting goals involved in the problem, since waste bins are considered {semi}-obnoxious facilities~\citep{tralhaoX2010multiobjective}. On the one hand, citizens that live near to the bins can suffer different environmental costs, such as noise pollution, bad smell, visual pollution, and traffic congestion from collection vehicles~\citep{bautista2006modeling,flahautX2002locating,khan2016allocation}. On the other hand, citizens that live far from the bins can suffer large transportation costs, having to carry their waste for long distances, which can affect the accessibility of the system. Citizens unwilling to incur in this transportation might dump their waste in unsuitable places~\citep{parrotX2009municipal,sotamenou2019drivers}. Dumped waste must be removed by the authorities, incurring in additional expenses. This conflicting relation between the environmental and transportation costs is associated with the NIMBY response to undesirable facilities: few citizens accept these facilities to be placed nearby, compared to the number of citizens who admit that they should be placed somewhere else~\citep{lindell1983close}. Related to this, \citet{ghianiX2012capacitated} and \citet{singh2019managing} stated that despite their eagerness to have a bin as close as possible, citizens are also willing to have the smaller tax burden that guarantee the service, which affects the budget to install several bins in an urban area. 

In turn, decision-makers should avoid locating waste bins near water streams or certain public places, such as schools, hospitals, and religious buildings~\citep{ahmed2006solid,ugwuishiwu2020gis}. This consideration is not only related to the fact that bins may contribute to visual pollution of an urban area, which can even affect the commercial selling price of the nearby buildings~\citep{di2014integrationpilotcase}, but also an contribute to the dissemination of vector-borne diseases~\citep{ali2019spatial,gupta2019use}. In this regard, \citet{nesmachnow2018comparison} aimed at providing a frequent collection service to those waste bins located near busy places, i.e., assigning them a higher priority in the collection schedule, so they are unlikely to be overflowed.

\subsection{\blue{Relevant considerations of related works}}
Sections~\ref{sec:objective_criteria}--\ref{sec:integrated} present the main details of the literature review on the waste bins location problem. Relevant issues are described and discussed, including the applied optimization criteria, resolution methods, and integrated approaches with other stages of the reverse MSW supply chain. 
The main details of the reviewed articles are reported in Table~\ref{tab:criteria} (optimization criteria and resolution approaches) and Table~\ref{tab:scenarios} (case studies and main results). Table~\ref{tab:scenarios}, categorizes the articles considering the type of waste: only installing bins for the collection of recyclable materials, installing bins for both recyclable and non-recyclable material, and installing bins for a unique stream of mixed waste.

{\scriptsize
\begin{longtable}{lccccl}
\caption{Waste bins location problem: optimization criteria and resolution methods.}\\
\label{tab:criteria}\\[-7pt]
\setlength{\tabcolsep}{3pt}
\renewcommand{\arraystretch}{1.1}
\setlength\LTleft{75pt}           
\setlength\LTright{100pt}   
\\[-7pt]\hline\\[-7pt]
\multicolumn{1}{c}{\multirow{2}{*}{\textit{Article}}} & \multicolumn{4}{c}{\textit{Optimization criteria}} & \multicolumn{1}{c}{\multirow{2}{*}{\textit{Resolution method}}} \\
\cline{2-5}\\[-7pt]
& C & UA & RC & O \\
\\[-7pt]\hline\\[-7pt]
\endfirsthead
\multicolumn{6}{l}{\tablename\ \thetable\ -- \textit{Continued from previous page}} \\
\\[-7pt]\hline\\[-7pt]
\multicolumn{1}{c}{\multirow{2}{*}{\textit{Article}}} & \multicolumn{4}{c}{\textit{Optimization criteria}} & \multicolumn{1}{c}{\multirow{2}{*}{\textit{Resolution method}}} \\
\cline{2-5}\\[-7pt]
& C & UA & RC & O \\
\\[-7pt]\hline\\[-7pt]
\endhead
\\[-7pt]\hline\\[-7pt] \multicolumn{6}{r}{\textit{Continued on next page}} \\
\endfoot
\\[-7pt]\hline\\[-7pt]
\endlastfoot
\citet{adedotun2020improving} &  & $\times$ & $\times$ &  & Manual location with the aid of GIS-based information \\ 
\citet{adeleke2021efficient} & $\times$ & $\times$ &  &  & Linear Lagrangian relaxation heuristic \\ 
\citet{ahmed2006solid} &  & $\times$ &  & $\times$ & Manual location with the aid of GIS-based information \\ 
\citet{aka2018fuzzy} & $\times$ & $\times$ &  &  & Fuzzy Multiobjective MILP \\ 
\citet{aremu2012case} &  & $\times$ & $\times$ & $\times$ & Single-objective MILP ($p$-median model inside GIS) \\ 
\citet{aremu2012framework} & $\times$ & $\times$ & $\times$ & $\times$ & Single-objective MILP ($p$-median model inside GIS) and AHP \\ 
\citet{aremu2016modeling} &  & $\times$ &  &  & ArcGIS Network Analyst \\ 
\citet{barrena2019optimizing} & $\times$ &  &  &  & Greedy algorithm \\ 
\citet{barrena2020solidarity} & $\times$ & $\times$ &  &  & Greedy algorithm with later improvement steps \\ 
\citet{bautista2006modeling} & $\times$ & $\times$ &  &  & Single-objective MILP \\ 
\citet{bennekrouf2020strategic} & $\times$ & $\times$ &  &  & Single-objective MILP \\ 
\citet{blazquez2020network} & $\times$ & $\times$ & $\times$ &  & Single-objective MILP \\ 
\citet{boskovic2015fast} &  & $\times$ & $\times$ &  & Manual location with the aid of GIS-based information \\ 
\citet{carlos2016optimization} & $\times$ & $\times$ &  &  & Manual location with the aid of GIS-based information \\ 
\citet{carlos2021design} &  & $\times$ &  &  & Manual location with the aid of GIS-based information \\ 
\citet{cavallin2020application} & $\times$ & $\times$ &  &  & Single-objective MILP \\ 
\citet{chang1999strategic} &  & $\times$ & $\times$ & & Evolutionary algorithm \\
\citet{chang2000siting} &  & $\times$ & $\times$ & & Fuzzy evolutionary algorithm \\ 
\citet{coutinhoX2012bi} & $\times$ & $\times$ &  & $\times$ & Multiobjective MILP \\ 
\citet{cubillos2020solution} & & $\times$ & $\times$ & & Multiobjective MILP and Variable Neighborhood Search Heuristic \\
\citet{di2014integration,di2014integrationpilotcase} &  & $\times$ &  &  & Constructive heuristic \\ 
\citet{erfani2017novel} & $\times$ & $\times$ &  &  & ArcGIS Network Analyst \\ 
\citet{erfaniX2018using} &  & $\times$ & $\times$ &  & ArcGIS Network Analyst \\ 
\citet{ferronato2020assessment} & & $\times$ & & & Location method not specified (using GIS information) \\
\citet{flahautX2002locating} &  & $\times$ &  & $\times$ & Single-objective MILP ($p$-median model) \\ 
\citet{gallardoX2015methodology} & $\times$ & $\times$ &  &  & ArcGIS Network Analyst \\ 
\citet{gautam2005strategic} &  & $\times$ &  &  & Single-objective MILP ($p$-median model) \\ 
\citet{ghianiX2012capacitated} & $\times$ & $\times$ &  &  & Single-objective MILP and Constructive heuristic \\ 
\citet{ghianiX2014impact} & $\times$ & $\times$ & $\times$ &  & Single-objective MILP and Constructive heuristic \\ 
\citet{gilardino2017combining} & $\times$ & $\times$ &  &  & Single-objective MILP \\ 
\citet{gonzalez2002model} &  &  & $\times$ & $\times$ & Single-objective MILP \\ 
\citet{hemmelmayrX2013models} & $\times$ &  & $\times$ &  & Single-objective MILP and Variable Neighborhood Search Heuristic \\ 
\citet{hemmelmayrX2017periodic} & $\times$ &  & $\times$ & $\times$ & Single-objective MILP and Adaptive Large Neighborhood Search \\ 
\citet{herrera2018optimization} & $\times$ & $\times$ &  &  & Multiobjective MILP \\ 
\citet{jammeliX2019bi} & $\times$ &  & $\times$ &  & Clustering heuristic and stochastic single-objective MILP \\ 
\citet{kaoX2002shortest} & $\times$ & $\times$ &  &  & Single-objective MILP \\
\citet{kao2013spatial} & $\times$ & $\times$ &  &  & Single-objective MILP \\ 
\citet{kao2010service} &  & $\times$ &  &  & Single-objective MILP \\ 
\citet{karadimasX2005gis} &  &  & $\times$ &  & Location method not specified (using GIS information) \\ 
\citet{karadimas2008gis} &  &  & $\times$ &  & Location method not specified (using GIS information) \\ 
\citet{karkanias2014assessing} &  & $\times$ &  & $\times$ & Location method not specified (using GIS information) \\ 
\citet{khan2016allocation} &  & $\times$ &  &  & Manual location \\ 
\citet{kim2013restricted} & $\times$ & $\times$ &  &  & Single-objective MILP, multi-stage Branch and Bound, and drop heuristic \\ 
\citet{kim2015case,kim2015integrated} & $\times$ &  & $\times$ &  & Tabu Search \\ 
\citet{letelier2021solving} & $\times$ & $\times$ &  &  & Single-objective MILP for non-recyclable bins and ArcGIS for recyclable bins \\ 
\citet{linX2011model} & & $\times$ & $\times$ & & Single-objective MILP \\
\citet{lopez2008optimizing} & & $\times$ & $\times$ & $\times$ & Manual location with the aid of GIS-based information \\
\citet{lopez2009containerisation} & $\times$ & $\times$ &  &  & Manual location with the aid of GIS-based information \\ 
\citet{maraqa2018optimization} & $\times$ &  &  &  & Manual location \\ 
\citet{nevrly2019municipal} & $\times$ & $\times$ &  &  & Multiobjective MILP \\ 
\citet{nevrly2021location} & $\times$ & $\times$ & $\times$ &  & Multiobjective MILP \\ 
\citet{nithya2012optimal} & $\times$ &  &  & & Manual location with the aid of GIS-based information \\ 
\citet{oliaeibreakdown} &  &  &  & $\times$ & Location method not specified (using GIS information) \\ 
\citet{paul2017using} &  & $\times$ &  & & Manual location with the aid of GIS-based information \\ 
\citet{rathore2019allocation} & $\times$ &  &  &  & Single-objective MILP and ArcGIS  \\ 
\citet{rathore2020location} & $\times$ &  &  &  & Single-objective MILP and ArcGIS  \\ 
\citet{ratkovic2016planning} & $\times$ & $\times$ &  &  & Single-objective MILP \\ 
\citet{rossitX2017application} & $\times$ & $\times$ &  &  & Multiobjective MILP \\ 
\citet{rossit2018municipal} & $\times$ & $\times$ & $\times$ &  & Multiobjective MILP \\ 
\citet{rossit2019bi} & $\times$ & $\times$ & $\times$ &  & Multiobjective MILP \\ 
\citet{rossit2020exact} & $\times$ & $\times$ & $\times$ &  & Multiobjective MILP and PageRank \\ 
\citet{sheriff2017integrated} & $\times$ & $\times$ & $\times$ & $\times$ & Single-objective MILP \\ 
\citet{toutouh2018intelligence} & $\times$ & $\times$ &  & $\times$ & Multiobjective evolutionary algorithms and PageRank \\ 
\citet{toutouh2020soft} & $\times$ & $\times$ &  & $\times$ & Multiobjective evolutionary algorithms and PageRank \\ 
\citet{tralhaoX2010multiobjective} & $\times$ & $\times$ &  & $\times$ & Multiobjective MILP \\ 
\citet{ugwuishiwu2020gis} &  &  &  & $\times$ & Manual location with the aid of GIS-based information \\
\citet{valeo1998location} & $\times$ & $\times$ &  &  & ArcGIS Network Analyst \\ 
\citet{vidovic2016two} & $\times$ & $\times$  &  & $\times$ & Single-objective MILP and Two-phase (greedy + LP) heuristic \\ 
\citet{vijay2005estimation} &  & $\times$ & &  & Greedy heuristic \\ 
\citet{vijay2008gis} &  & $\times$ &  $\times$ &  & Single-objective MILP ($p$-median model) \\ 
\citet{vu2018parameter} & $\times$ & $\times$ & & $\times$ & ArcGIS Network Analyst \\ 
\citet{yaakoubi2018heuristic} &  &  & $\times$ &  & Single-objective MILP and a two-phase (Memetic algorithm + ILS) heuristic  \\ 
\citet{zahan2020multi} &  &  & $\times$ & $\times$ & AHP \\ 
\citet{zamorano2009planning} & $\times$ & $\times$ & $\times$ &  & ArcGIS Network Analyst \\ 
\midrule
\multicolumn{6}{c}{\multirow{2}{\textwidth}{Abbreviations used: AHP: Analytical Hierarchy Procedure; GIS: Geographic Information System; ILS: Iterated Local Search\\
LP: Linear Programming; MILP: Mixed Integer Linear Programming.}} \\
\\
\end{longtable}
}

{\scriptsize
\begin{longtable}{lp{2.75cm}p{9cm}}
\caption{Waste bins location problem: case studies and main results.}
\label{tab:scenarios}\\[-7pt]
\setlength{\tabcolsep}{0pt}
\setlength\LTleft{125pt}           
\setlength\LTright{100pt}   
\renewcommand{\arraystretch}{1.1}
\\[-7pt]\hline\\[-7pt]
\multicolumn{1}{c}{\textit{Article}} & \multicolumn{1}{c}{\textit{Case studies}} & 
\multicolumn{1}{c}{\textit{Main results}} \\
\\[-7pt]\hline\\[-7pt]
\endfirsthead
\multicolumn{3}{l}{\tablename\ \thetable\ -- \textit{Continued from previous page}} \\
\\[-7pt]\hline\\[-7pt]
\multicolumn{1}{c}{\textit{Article}} & 
\multicolumn{1}{c}{\textit{Case studies}} &
\multicolumn{1}{c}{\textit{Main results}} \\
\\[-7pt]\hline\\[-7pt]
\endhead
\\[-7pt]\hline\\[-7pt] \multicolumn{3}{r}{\textit{Continued on next page}} \\
\endfoot
\\[-7pt]\hline\\[-7pt]
\endlastfoot
\multicolumn{3}{c}{Installation of bins for the collection of recyclable materials} \\
\\[-7pt]\hline\\[-7pt]
\citet{aka2018fuzzy} & Antalya\,(TR) & Compromise solutions (recyclable material collected and collection routing distance) \\
\citet{bennekrouf2020strategic} & Boudjlida\,(DZ) & Design of a recyclable material collection network of bins considering QoS and investment cost \\ 
\citet{chang1999strategic} & Kaohsiung\,(TW) & Set of compromising solutions between QoS, routing cost, and population served by the system for recyclable material  \\ 
\citet{chang2000siting} & Kaohsiung\,(TW) & Set of compromising solutions between QoS, routing cost, and population served by the system for recyclable material \\ 
\citet{cubillos2020solution} &  Five Danish cities & Set of compromising solutions between the population served and the routing costs for recyclable material \\ 
\citet{flahautX2002locating} & La Buyère\,(BE) & Compromising solution regarding QoS and environmental costs \\ 
\citet{gautam2005strategic} & No real case solved & Methodology to find solutions to maximize the QoS \\ 
\citet{gonzalez2002model} & Asturias\,(ES) & Increment of the amount of recyclable material collected \\ 
\citet{hemmelmayrX2017periodic} & Not specified region & Reduction of total costs when flexible location of vehicle depots or visit schedules for collection points are considered \\
\citet{kao2010service,kao2013spatial} & Hsinchu\,(TW) & Solutions with high QoS for collecting recyclable material \\ 
\citet{linX2011model} & Taichung\,(TW) & Methodology for designing a two-shift bins collection network for recyclable waste  \\ 
\citet{lopez2008optimizing} & Madrid\,(ES) & Improved efficiency of the collection system by designing a separated collection system for paper/cardboard of small businesses \\ 
\citet{lopez2009containerisation} & Aranjuez\,(ES) & Reduction of the number of collection points, improved network coverage and increased  collected recyclable material \\ 
\citet{ratkovic2016planning} & Belgrade\,(RS) & Methodology for designing a recyclable waste network of bins \\ 
\citet{sheriff2017integrated} & Not\,specified\,town\,(IN) & Integrated approach solving bins location and collection routing outperformed sequential approaches for recyclable material \\ 
\citet{valeo1998location} & Dundas\,(CA) & Methodology for recyclable bins network design \\
\citet{vidovic2016two} & No real case study solved & Methodology for designing the network of bins, collection routes, and transfer stations of recyclable material \\
\\[-7pt]\hline\\[-7pt]
\multicolumn{3}{c}{Installation of bins for both recyclable and non-recyclable material} \\
\\[-7pt]\hline\\[-7pt]
\citet{adeleke2021efficient} & Lagos\,(NG) & Reduction of activated collection sites and allocated bins \\
\citet{ahmed2006solid} & Aurangabad\,(IN) & Proposal for a source classified bins network \\
\citet{barrena2019optimizing,barrena2020solidarity} & Seville\,(ES) & Reduction of bin network costs considering a solidarity behavior of citizens \\ 
\citet{carlos2016optimization} & Castellón\,(ES) & Solution with improved QoS \\ 
\citet{carlos2021design} & Nikki\,(BJ) & Methodology for enhancing formal waste collection \\ 
\citet{cavallin2020application} & Bahía Blanca\,(AR) & Methodology for migrating from a house-to-house collection a community bins-based collection \\ 
\citet{coutinhoX2012bi} & Coimbra\,(PT) & Set of compromising solutions considering QoS, investment cost, and semi-obnoxiousness of the bins \\ 
\citet{di2014integration,di2014integrationpilotcase} & L'Aquila\,(IT) & Set of solutions for different levels of QoS \\ 
\citet{ferronato2020assessment} & La Paz\,(BO) & Reduction of cost of the network of bins considering informal recycling sector \\
\citet{gallardoX2015methodology} & Castellón\,(ES) & Methodology to design a MSW management plan depending on the available data \\ 
\citet{gilardino2017combining} & Lima\,(PE) & Reduction of the routing costs in comparison to kerbside collection \\ 
\citet{karkanias2014assessing} & Neapoli-Sykies\,(GR) & Bins redistribution enhanced recycling rates and investment cost \\ 
\citet{letelier2021solving} & Santiago\,(CL) & QoS improvements for recyclable waste bins \\ \citet{oliaeibreakdown} & Tabriz\,(IN) & Methodology to locate bins \\ 
\citet{rathore2019allocation} & Bilaspur\,(IN) & Reduction of the number of bins \\ 
\citet{rathore2020location} & Bilaspur\,(IN) & Reduction of the number of bins that led to a reduction of idling cost and carbon emissions of collection vehicles \\
\citet{rossitX2017application} & Bahía\,Blanca\,(AR) & Set of compromise solutions between installment costs and QoS \\
\citet{rossit2019bi} & Bahía\,Blanca\,(AR) & Set of compromise solutions between installment costs and collection frequency \\ 
\citet{rossit2020exact} & Montevideo\,(UY), Bahía\,Blanca\,(AR) & Set of compromise solutions regarding collection frequency, installment cost, and QoS \\ 
\citet{tralhaoX2010multiobjective} & Coimbra\,(PT) & Compromise solutions regarding QoS, investment cost, and semi-obnoxiousness \\ 
\citet{zamorano2009planning} & Granada\,(ES) & Reduction of the number of bins and collection costs \\  
\\[-7pt]\hline\\[-7pt]
\multicolumn{3}{c}{Installation of bins for a unique stream of mixed waste} \\
\\[-7pt]\hline\\[-7pt]
\citet{adedotun2020improving} & Ibadan\,(NG) &  Reduction of the collection costs \\  
\citet{aremu2012case} & Ilorin\,(NG) & Set of compromising solutions between QoS to citizens, collection routing criteria, and investment costs \\ 
\citet{aremu2012framework} & Ilorin\,(NG) & Compromising solution regarding social, economic, and environmental criteria \\ 
\citet{aremu2016modeling} & Ilorin\,(NG) & Compromising solutions between QoS and investment costs \\  
\citet{bautista2006modeling} & Barcelona\,(ES) & Algorithms to aid decision-makers for locating bins \\ 
\citet{blazquez2020network} & Santiago\,(CL) & Compromising solutions between QoS and investment costs. \\ 
\citet{boskovic2015fast} & Kragujevaco\,(RS) & Reduction of the number of collection points and number of bins \\ 
\citet{erfani2017novel} & Mashhad\,(IR) & Improved bin distribution that led to reduced routing costs \\
\citet{erfaniX2018using} & Mashhad\,(IR) & Set of compromising solutions between QoS, area coverage, population served, unused capacity of bins, and the balanced distribution of waste among all the installed bins \\ 
\citet{ghianiX2012capacitated} & Nardò\,(IT) & Reduction of the number of collection points and the number of bins \\ 
\citet{ghianiX2014impact} & Nardò\,(IT) & Improved bin distribution that led to reduced routing costs \\ 
\citet{hemmelmayrX2013models} & {Not specified town in Northern IT} & Integrated approach outperformed hierarchical approaches in terms of joint cost of bins allocation and routing \\ 
\citet{herrera2018optimization} & Ibarra\,(EC) & Design of a collection network of bins considering QoS and investment cost  \\ 
\citet{jammeliX2019bi} & Sousse\,(TN) & Reduction of the combined investment cost of routing and bins location considering stochastic waste generation \\ 
\citet{kaoX2002shortest} & Hsinchu\,(TW) & Solutions with high QoS \\
\citet{karadimasX2005gis} & Athens\,(GR) & Reduction of the number of bins required \\ 
\citet{karadimas2008gis} & Athens\,(GR) & Reduction of the number of bins \\ 
\citet{khan2016allocation} & Dhanbad\,(IN) & Methodology for optimizing bin locations \\ 
\citet{kim2013restricted} & No real case study solved & Methodology for optimizing bin locations considering fluctuating waste generation \\ 
\citet{kim2015case,kim2015integrated} & Seoul\,(KR) & Integrated approach (bins location and collection routing) outperformed sequential methods for fluctuating waste generation \\ 
\citet{maraqa2018optimization} & Al Ain city\,(AE) & Reduction of the number of bins, and fuel consumption/pollutants emissions of the collection vehicle \\ 
\citet{nevrly2019municipal} & Tábor\,(CZ) & Methodology for locating bins with minimal utilization rate \\ 
\citet{nevrly2021location} & Tábor\,(CZ) & Extensive analysis of the trade-off among diverse objectives \\ 
\citet{nithya2012optimal} & Coimbatore\,(IN) & Increment of the area coverage of the network of bins \\
\citet{paul2017using} & Kolkata\,(IN) & Redistribution of bins increment of the total area serve by the network of bins \\ 
\citet{rossit2018municipal} & Bahía\,Blanca\,(AR) & Set of compromise solutions between installment costs and collection frequency \\ 
\citet{toutouh2018intelligence} & Montevideo\,(UY) & Set of compromise solutions between population served, QoS, and installment costs \\ 
\citet{toutouh2020soft} & Montevideo\,(UY), Bahía\,Blanca\,(AR) & Set of compromise solutions between population served, QoS, and installment costs outperformed current situation in terms of QoS. \\ 
\citet{ugwuishiwu2020gis} & Enugu\,(NG) & Improved distribution of bins avoiding busy/inconvenient places \\ 
\citet{vijay2005estimation} & No real case study solved & Methodology for estimating waste generation and allocating users to bins imporving the QoS provided to the users \\ 
\citet{vijay2008gis} & No real case study solved & Methodology for estimating waste generation and allocating users to bins imporving the QoS provided to the users \\
\citet{vu2018parameter} & Hai Phong\,(TW) & Solutions for different QoS and total number of bins to be located \\ 
\citet{yaakoubi2018heuristic} & No real case study solved & Methodology to solve sequentially the problems of locating bins and scheduling collection routes \\ 
\citet{zahan2020multi} & Dhaka\,(BD) & Multi-attribute methodology to choose locations for bins considering QoS and environmental issues \\ 
\end{longtable}
}

\section{\blue{A quantitative analysis of literature}}
\label{sec:quantitative}

This section presents a quantitative analysis of the 76 reviewed related works.

\subsection{Bibliometric analysis} 

Of the 76 revised articles, 16 were published in conferences and 60 in journals. 
These numbers suggest that researchers consider the subject relevant enough to submit to indexed journals rather than conferences. In turn, {18 articles were published before 2011 and 58 in the last ten years}. Figure~\ref{Fig:pub_by_year} reports the chronological distribution of conference and journal articles (number of publications per year and per type). Articles published before 2008 are grouped together in the first columns. The last column groups articles published in 2010 and in the first three months of 2021.

\pgfplotsset{
    axis line style={black},
    every axis label/.append style ={black},
    every tick label/.append style={black} 
}

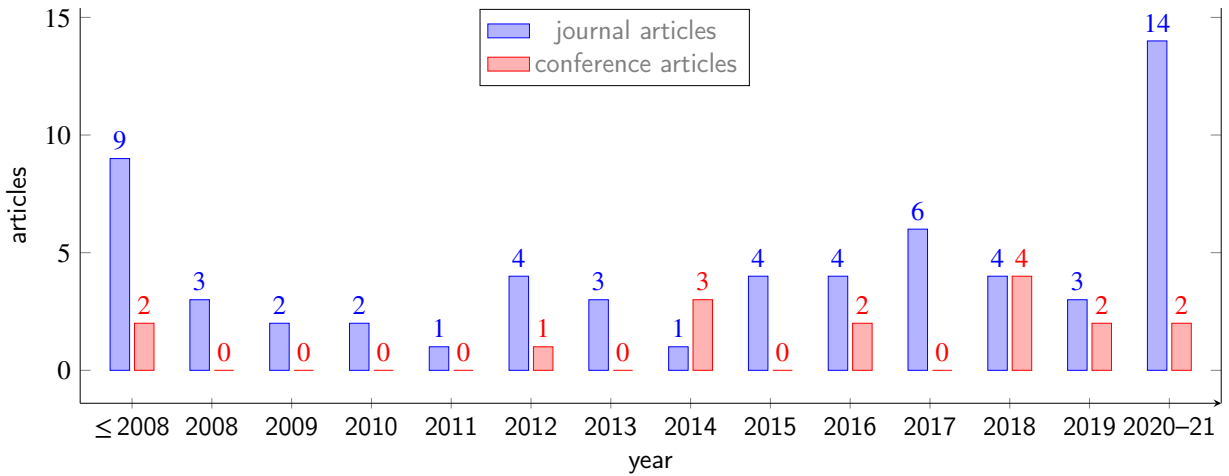
\begin{figure}[!h]
\setlength{\abovecaptionskip}{-6pt}
\resizebox{0.99\textwidth}{!}{%
\begin{tikzpicture}
\begin{axis}[
    every axis plot post/.style={/pgf/number format/fixed},
    ybar = 2pt, area legend,
    bar width = 8pt,
    x = 1.15cm,
    symbolic x coords={$\le$\,2008, 2008, 2009, 2010, 2011, 2012, 2013, 2014, 2015, 2016, 2017, 2018, 2019, 2020--21}, 
    legend style={at={(0.35,0.9)},anchor=west},
    axis x line = bottom, nodes near coords, 
    ylabel = articles,
    xlabel = year,
    enlarge x limits=0.05, 
    ] 
]
\addplot+ coordinates {($\le$\,2008, 9) (2008, 3) (2009, 2) (2010,2) (2011, 1) (2012, 4) (2013, 3) (2014, 1) (2015,4) (2016, 4) (2017, 6) (2018, 4) (2019, 3) (2020--21,14)}; 
\addplot+ coordinates {($\le$\,2008, 2) (2008, 0) (2009, 0) (2010,0) (2011, 0) (2012, 1) (2013, 0) (2014, 3) (2015,0) (2016, 2) (2017, 0) (2018, 4) (2019, 2) (2020--21, 2)}; 
\legend{journal articles,conference articles}; 
\end{axis} 
\end{tikzpicture} 
}
\vspace{-0.3cm}
\caption{Journal and conference publications related to the waste bins location problem, by year}
\label{Fig:pub_by_year}
\end{figure}

Values reported in Figure~\ref{Fig:pub_by_year} show a renewed interest on the bin location problem, as demonstrated by the 16 recent articles on that subject (12 articles published in 2020 and 4 articles published in just three months of 2021). Figure~\ref{Fig:journals} presents the most representative journals that have published articles related to the waste bins location problem: {Waste Management} (Elsevier) with 10 articles, {Waste Management \& Research} (SAGE Publications) with 8 articles; {Journal of Environmental Engineering} (American Society of Civil Engineers) and \textit{Journal of Cleaner Production} (Elsevier) both with 3 articles; and, finally, {International Journal of Industrial and the Systems Engineering} (Inderscience Enterprises Ltd.), {Journal of Operational Research Society} (Taylor and Francis Ltd.) and {Journal of the Air \& Waste Management Association} (Taylor and Francis Ltd.) with 2 articles. The remaining 38 articles were published in other general-purpose journals.

\definecolor{color1}{HTML}{D7191C}
\definecolor{color2}{HTML}{FDAE61}
\definecolor{color3}{HTML}{ABDDA4}
\definecolor{color4}{HTML}{2B83BA}
\definecolor{color5}{HTML}{6A2A1C}
\definecolor{color6}{HTML}{DDE16C}
\definecolor{color7}{HTML}{54985E}
\begin{figure}[!h]
\setlength{\abovecaptionskip}{3pt}
\centering
\resizebox{0.975\textwidth}{!}{%
\begin{tikzpicture}
\begin{axis}[
    xbar stacked,
    legend style={legend columns=4, at={(xticklabel cs:0.5)}, anchor=north,  draw=none},
    ytick=data,
    axis y line*=none,
    axis x line*=bottom,
    tick label style={font=\footnotesize},
    legend style={font=\footnotesize},
    label style={font=\footnotesize},
    xtick={0,5,10,15},
    width=.6\textwidth,
    bar width=6mm,
    yticklabels={
     {Journal of the Air \& Waste Management Association}, 
    {Journal of the Operational Research Society}, {Int.~Journal of Industrial and Systems Engineering}, {Journal of Cleaner Production}, {Journal of Environmental Engineering}, {Waste Management and Research},{Waste Management}},
    xmin=0, xmax=12,
    area legend,
    y=8mm,
    enlarge y limits={abs=0.625},
    nodes near coords, nodes near coords style={text=black, at ={(\pgfplotspointmeta,\pgfplotspointy)},anchor=west},
    visualization depends on=y \as \pgfplotspointy,
    every axis plot/.append style={fill}
]
\addplot[color1] coordinates
  {
  (0,0) 
  (0,1) (0,2) (0,3) (0,4) (0,5) (10,6)};
\addplot[color2] coordinates
  {
  (0,0) 
  (0,1) (0,2) (0,3) (0,4) (8,5) (0,6)};
\addplot[color3] coordinates
  {
  (0,0) 
  (0,1) (0,2) (0,3) (3,4) (0,5) (0,6)};
\addplot[color4] coordinates
  {
  (0,0) 
  (0,1) (0,2) (3,3) (0,4) (0,5) (0,6)};
\addplot[color5] coordinates
  {
  (0,0) 
  (0,1) (2,2) (0,3) (0,4) (0,5) (0,6)};
\addplot[color6] coordinates
  {
  (0,0) 
  (2,1) (0,2) (0,3) (0,4) (0,5) (0,6)};
  \addplot[color7] coordinates
  {(2,0) (0,1) (0,2) (0,3) (0,4) (0,5) (0,6)};
\end{axis}  
\end{tikzpicture}
}
\vspace{-0.3cm}
\caption{
{Most representative journals where the related works were published.}}
\label{Fig:journals}
\end{figure}
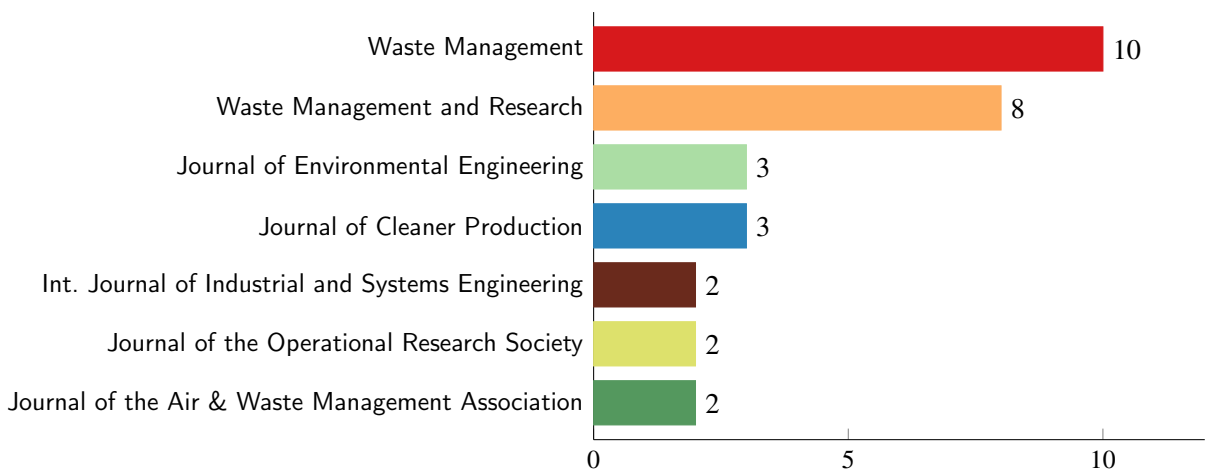

\subsection{Analysis of the addressed scenarios}

Another relevant aspect to analyze is the scenarios where the methodologies proposed in related works were applied and evaluated. Two common dimensions from the 
related literature are analyzed: the use of real-world case studies and the consideration of separated recyclable materials. 

\paragraph{Use of real-world case studies.} 
Case studies based on real-world data have been relevant for the evaluation of methodologies proposed to solve the waste bins location problem. 88\% of the related works (67 out of 76) have solved scenario(s) considering real data from at least one city (Table~\ref{tab:scenarios}). In general, the real information used were GIS maps and waste generation rates inferred from algebraic formulas based on population distribution and commercial activities. Getting real waste generation data is a time consuming process and the obtained information rapidly outdates, due to its great variability~\citep{lebersorger2011municipal}. \blue{For example, the scarcity of accurate data has been identified as the main drawback for implementing efficient collection systems in developing countries~\citep{carlos2021design}.
}
However, real information is important to test the real applicability of the proposed models and is mandatory to guarantee replicability. Articles using real waste generation data have used information obtained either from surveys or analysis performed by the authors~\citep{cavallin2020application,karadimas2008gis}, from other private sources~\citep{maraqa2018optimization,nevrly2021location}, or from studies performed in other similar cities~\citep{cavallin2020application,carlos2021design}. Some articles used public repositories~\citep{bautista2006modeling,ferronato2020assessment,ratkovic2016planning,rossit2020exact} and a few works have used (non-realistic) randomly generated data for waste generation rate~\citep{adeleke2021efficient,barrena2019optimizing,barrena2020solidarity}. Finally, some articles have located uncapacitated collection points~\citep{aka2018fuzzy,kaoX2002shortest,valeo1998location}, without needing
to estimate waste generation rate
but only providing a partial (unrealistic) view of the problem. 

Most articles that considered the capacity of bins used realistic commercial bins. A few works used unreal capacities~\citep{adeleke2021efficient,rossitX2017application}. Some articles have estimated the available surface space to locate bins~\citep{hemmelmayrX2013models,cavallin2020application,rossit2020exact}, taken from real data~\citep{ghianiX2012capacitated,ghianiX2014impact} or randomly generated~\citep{barrena2019optimizing,barrena2020solidarity}.
A digital map of the scenario is another relevant input. Some articles developed their own digital maps~\citep{boskovic2015fast,carlos2021design,chang1999strategic,chang2000siting} or obtained them from specific databases of public institutions~\citep{ahmed2006solid,valeo1998location}, whereas many recent articles took advantage from the online availability of real maps, such as OpenStreetMap~\citep{rossit2020exact,toutouh2020soft}, Google Earth~\citep{aremu2016modeling,khan2016allocation}, or Google Maps~\citep{adeleke2021efficient}. Maps have been usually processed using GIS software, to visualize geographic information and calculate urban walking distances. \citet{tralhaoX2010multiobjective} developed their own GIS application. 
A few works, although claim to use real data, do not made clear the data sources~\citep{nevrly2019municipal,zahan2020multi}.

Three articles considered case studies from more than one city: \citet{rossit2020exact} and~\citet{toutouh2020soft} solved case studies from Bahía Blanca, Argentina and Montevideo, Uruguay, whereas \citet{cubillos2020solution} solved case studies from five different Danish cities.

\begin{figure}[!h]
\centering
\begin{minipage}{.49\linewidth}
  	\includegraphics[width=1\textwidth]{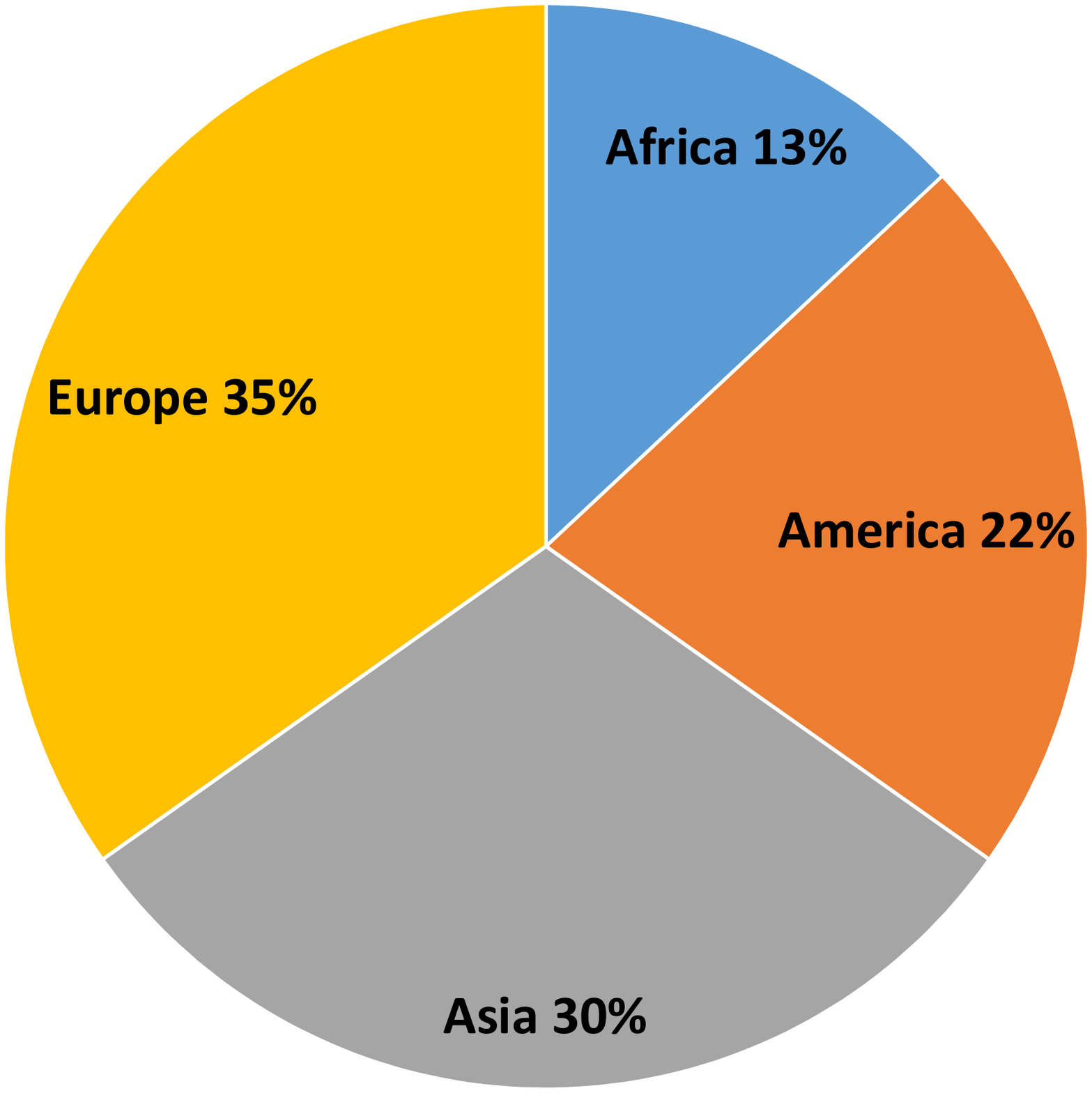}
  \end{minipage}
  \begin{minipage}{.49\linewidth}
    \includegraphics[width=1\textwidth]{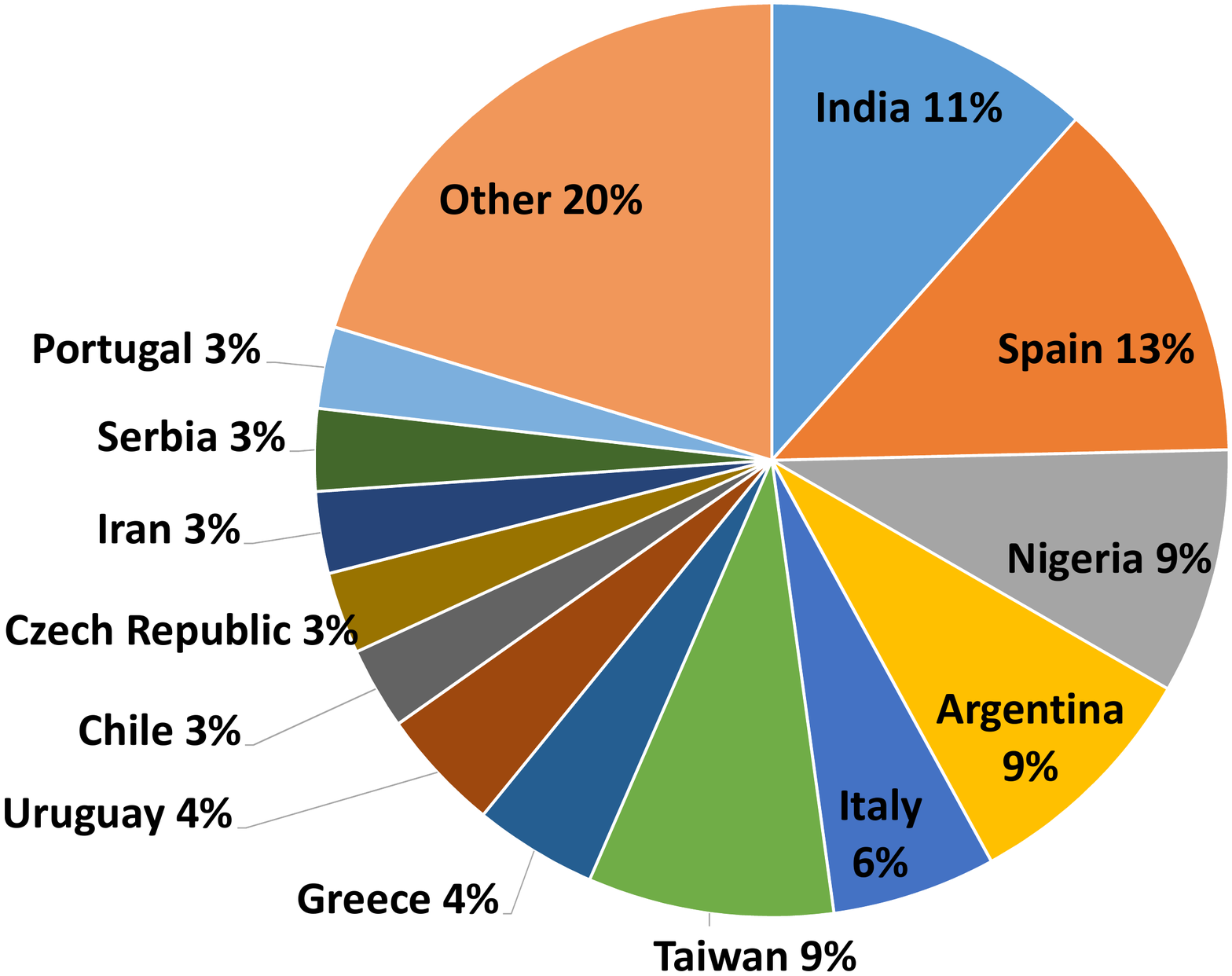}
  \end{minipage}
  \vspace{-0.2cm}
  \caption{Articles that solved real-world case studies from the related literature classified by continent (left) and country (right).\label{fig:case_studies}}
\end{figure}

Figure~\ref{fig:case_studies} presents the distribution of case studies per continent and country. Regarding continents, most case studies were located in Europe (24) and Asia (21). Regarding countries, the most represented were Spain (9 works), India (8 works), Taiwan, Argentina, and Nigeria (6 works).

\vspace{-2mm}
\paragraph{Consideration of separated recyclable material collection.} 
A good planning of waste bins location 
can largely contribute to the success of recycling programs. In this context, 23\% of the reviewed articles (18 out of 76) have solved the waste bins location problem for scenarios that consider only installing bins for the collection of recyclable materials. In turn, 29\% of the articles (22 out of 76) have considered at least one scenario with source classification, i.e., installing both bins for recyclable material and bins for non-recyclable material. Finally, 5\% of the articles (4 out of 76) discussed extensions
to scenarios considering recyclable materials and source classification of waste~\citep{bautista2006modeling,boskovic2015fast,nevrly2019municipal,nithya2012optimal}. The rest of the articles (32 out of 76) did not consider separated recyclable material, dealing with a unique stream of mixed waste instead.
\blue{
Recyclable materials are usually not biodegradable or putrescible; thus, they admit to be collected less often than 
organic or moisture waste~\citep{cavallin2020application}, specially in areas with warm and humid weather~\citep{carlos2021design}.
However, reducing too much the frequency collection of recyclable material is not advisable, because households/building do not have enough space to properly store classified waste~\citep{barrena2019optimizing,gallardoX2015methodology}, which requires a larger number of bins than unclassified waste. 

The related literature has acknowledged the difficulty of implementing selective waste collection in developing countries due to the lack of not only technical and budgetary resources~\citep{bennekrouf2020strategic}, but also environmental regulations~\citep{carlos2021design}. Another aspect that should be taken into account when designing the waste collection network in developing countries is the participation of informal workers. The activities of the informal sector reduces the amount of recyclable material to be collected by the formal MSW system which can affect the outcomes of the system.
\citet{ferronato2020assessment} found that the inclusion of informal workers allowed reducing the waste system expenses in around 10\%, increasing the recycling rate in 3.5\%, and reducing the distances traveled by compactor trucks in around 7\%.
%
Similar results were obtained in both studies, i.e., an 18\% and 22\% increase in the number of bins when considering selective waste collection, compared with the scenario with unclassified waste collection.
}

\subsection{Analysis of the pursued criteria}

From the information in Table~\ref{tab:criteria}, it is inferred that accessibility of users and the costs of the network of bins have been the most important criteria for locating bins. 76\% of the reviewed articles considered user accessibility criteria (58 out of 76). In turn, 59\% of the works (45 out of 76) considered the cost of the network of bins as an optimization criteria. Moreover, 46\% (35 out of 76) of the articles considered both criteria jointly in proposed models. The objective of routing cost has been also relevant, as 38\% of the works (29 out of 76) considered this criteria.

\subsection{Analysis of resolution approaches}

Different resolution approaches have been presented in the related literature. From Table~\ref{tab:criteria}, it is inferred that exact MILP models are predominant, as they have been applied in 50\% (38 out of 76) of the related works. Among them, 38\% (29 out of 76) used MILP as a unique resolution method, whereas 11\% (8 out of 76) applied MILP for comparing (and validating) heuristic approaches in small scenarios, and just one article used MILP as part of a multi-stage heuristic approach. 
whereas 11\% (8 out of 76) applied MILP for comparing (and validating) heuristic approaches in small scenarios and just one article used MILP as part of a multi-stage heuristic approach. In turn, 29\% of the articles (22 out of 76) used heuristic/metaheuristic approaches and 16\% (12 out of 76) used GIS for solving the waste bins location problem. Finally, 16\% of the articles (12 out of 76) locate bins manually and 7\% (5 out of 76) do not clearly outlined the procedure for locating bins.

Another important aspect in the resolution process is the consideration of uncertainty in the input parameters. In regard to waste generation rate, only 5\% of the articles (4 out of 76) considered this parameter as stochastic in their models. In turn, 8\% of the articles (6 out of 76) performed sensitivity analysis to test the proposed models for typical values of this parameter. Regarding the maximum threshold distance between users and assigned bins, a sensitivity analysis was performed in 18\% of the articles (14 out of 76). In regard to the maximum potential number of bins or collection points, 12\% (9 out of 76) applied sensitivity analysis. Then, 6\% (5 out of 76) performed sensitivity analysis over the collection frequency of bins. Finally, 5\% of the articles (4 out of 76) evaluated their models for different bins storage capacity.

\blue{Regarding bins storage capacity, \cite{adeleke2021efficient} found that 
increasing the capacity of bins by 40\% does not affect the number of 
active
collection points and 
affects the number of bins used in less than 10\% in all cases. However, the increment in the maximum walking distance between users and bins did had a great impact in the active collection points (an increment from 100m to 200m produces reductions up to 72\% in the number of active collection points).
\citet{carlos2021design} analyzed the required number of bins according to collection frequency. Diminishing the collection frequency from three times per week to once per week increases the number of required bins in 105\% for mixed waste and 34\% for organic waste. In a similar study, \citet{cavallin2020application} found than when increasing the collection frequency of recyclable material from three times per week to four times per week the number of bins was reduced up to 35\% for recyclable waste and 31\% for organic waste and the number of active collection points was reduced up to 31\%.

Two relevant relations have not been sufficiently studied in the related literature: 
the relation between collection frequency and the distance among bins, and 
the relation between distance 
 among bins and volume of bins.
\citet{blazquez2020network} performed sensitivity analysis 
comparing
the case in which the distance among collection sites (where bins are located) is required to be larger than 140m and the case in which 
no distance restriction is set. The case in which no restriction is set has about 49\% more active collection sites and 3\% more bins than the scenario with the restriction of 140m.
However, they do not solve the waste collection routing considering this variation, thus the relation between the distance among bins and collection frequency is not analyzed. \citet{blazquez2020network} also compared the outline of network of bins when only small volume bins are available, only larger volume bins are available, and when both types of bins are available. The scenario when large bins are available has 5\% less active collection sites and 23\% less bins than the scenario considering small bins. When both types of bins are considered, the collection network has 6\% less active collection sites and 12\% less bins than the scenario when only small bins are considered. However, they did not studied the effect of the variation of the volume of the bins in the distance among bins.
}

\subsection{Consideration of waste collection stage}

Besides the waste bins location problem, 26\% of the reviewed works (20 out of 76)
also addressed the waste collection stage of the MSW system. Particularly, 16\% (12 out of 76) used integrated approaches for locating waste bins and finding collection routes simultaneously. Finally, 17\% of the works (13 out of 76) presented sequential approaches, first solving the waste bins location problem and then, optimizing the waste collection routes over the defined network of bins.
%

\section{Optimization criteria}
\label{sec:objective_criteria}

This section analyzes the optimization criteria used in the related work about waste bins location. First, an overall description of the analysis and categorization is presented.
Then, a specific discussion is performed for each identified category.

\subsection{General discussion}

Different optimization criteria have been considered
for finding the best location of waste bins in urban areas. Most 
criteria are related to 
cost reduction and QoS improvement. 
Considered costs
include the installment and maintenance costs of the collection network (i.e., the bins distributed in the city). Operating costs are only considered in articles studying integrated models that include routing costs, salaries of the personnel, etc. Regarding QoS improvement, the main indicator is the accessibility of citizens to the system, i.e., minimize the average/total distance between users and their assigned bin, or assigning users to a bin that it is within a certain maximum threshold distance. Cost and QoS criteria result in conflicting objective functions for optimization. On the one hand, if the majority of users have a bin located close to their homes, the number of bins distributed in the city results to be very large, as 
studied in 
sensitivity analysis on the related literature~\citep{di2014integrationpilotcase,nevrly2021location,tralhaoX2010multiobjective}. In such cases, there is an associated incremental cost for purchasing and installing the bins. On the other hand, deploying a small number of bins or concentrating them in few points of the city reduce the cost, but also affect the accessibility to the system, due to the long distances users must transport their waste. 
Different alternatives have been explored
to deal with this trade-off,
including 
multiobjective approaches aimed at obtaining compromise solutions~\citep{coutinhoX2012bi,rossit2020exact,toutouh2020soft,tralhaoX2010multiobjective}
and optimizing
only one criteria in a single-objective model while restricting the other criteria to a certain threshold value~\citep{blazquez2020network,cavallin2020application}.

Besides the predominant applied criteria, specific works have 
considered other issues when solving the problem.
Since the configuration of the bins network collection has an impact on the posterior collection,
some
works included criteria aimed at bounding the costs of the 
posterior route collection. A common feature considered 
is the accessibility of the collection vehicles to the bins, avoiding locating bins in secondary streets with improper width or limiting the required collection frequency for emptying bins. A few authors have used other highly specific criteria, e.g., the avoidance of deploying bins near public places, such as schools or hospitals, where they might provoke a negative social reaction, or reducing the negative environmental impact of bins to households by considering the semi-obnoxiousness of these facilities. 

The considered criteria have been used in different ways in the proposed optimization procedures. Some authors used the criteria as objective functions in an formalized optimization tool, i.e., when a mathematical model is outlined and solved by a solver/algorithm, while other authors included the criteria as constraints, in the form that the final solution must respect a certain threshold value. Finally, a group of authors used the criteria as subjective goals in a more manual and less formalized optimization process, in which the location are decided by practitioners after the consultation of the information in a convenient format for the decision-making process. 

This review classifies the applied criteria in four categories: \textit{cost criteria}, for objectives that are related to the economics of the system; \textit{user accessibility criteria}, for objectives that aim at enhancing the ease for the citizens to drop their waste into the system; \textit{routing collection criteria} for objectives that are connected to reduce the complexity of the posterior collection costs; and \textit{other criteria} including all specific criteria that not belong to any of the previous classification. These categories are further described in the following subsections.

The information about the problem criteria and the resolution approaches applied to solve the waste bins location problem was presented in Table~\ref{tab:criteria}. The first column reports the bibliographic reference. Then, separate columns are included to account for optimization criteria: `C' stands for cost criteria, `UA' stands for user accessibility criteria, `RC' stands for routing collection criteria, and `O' stands for other criteria. The considered criteria are commented in the next subsection.

\subsection{Cost criteria}

Several works considered cost-based criteria when optimizing the location of waste bins. An expense that is usually included is the purchase cost of the bins, which varies according to the type of bin and  capacity~\citep{aremu2012framework,barrena2019optimizing,barrena2020solidarity,bennekrouf2020strategic,blazquez2020network,cavallin2020application,coutinhoX2012bi,hemmelmayrX2013models,herrera2018optimization,kim2015integrated,nevrly2021location,rathore2019allocation,rathore2020location,rossitX2017application,rossit2019bi,rossit2018municipal,rossit2020exact,toutouh2018intelligence,toutouh2020soft,tralhaoX2010multiobjective}. 
Some articles also considered the maintenance cost and a prorated purchase cost based on the lifetime of bins and the time horizon of the planning \citep{hemmelmayrX2013models,rossit2020exact,nevrly2021location}. 
This feature is important when comparing costs in different periods of time. For example, in works that addressed the integrated problem of locating bins and designing collection routes (see Section~\ref{sec:integrated}), travel distance costs are operational costs assigned on a daily basis, whereas the waste bin purchasing are strategic costs, since the service life of bins is assumed to be several years. Other articles also considered the set-up cost of preparing a special site to install the bins, either on the street or on the sidewalk, usually requiring the installation of signals, special painting, or basement construction \citep{kim2013restricted,kim2015integrated,ratkovic2016planning,vidovic2016two,sheriff2017integrated,cavallin2020application}. Moreover,~\citet{cavallin2020application} determined a subset of few \textit{ideal} sites among potential sites where bins can be installed.
Ideal sites are places conveniently located in the city that require a relatively small set-up cost (in comparison to non-ideal sites).
Thus, they are considered as priority by the optimization algorithm for bins location. Similarly, \citet{valeo1998location} considered a few of the available sites where the location of bins is mandatory for the optimization model. The rest of the sites are only used for bin location if required. Regarding other costs related to bins, \citet{rathore2020location} included the cost of waiting for unloading a bin by the collection vehicle, which varies according to the type of bin. \citet{tralhaoX2010multiobjective} and \citet{coutinhoX2012bi} included an extra cost for locating a special type of bin that requires a specific collection vehicle to be emptied. When addressing scenarios that already have an existing network of bins installed, \citet{hemmelmayrX2013models} estimated a cost for moving a bin from one place to another, adding to the model a reticence to move bins unless it is needed.

Closely related to the costs of the system, \citet{bautista2006modeling} minimized the total number of bins in a network (subject to fulfilling a required level of service), since it is directly related to the set-up cost of conditioning places for bins and to the visual and noise pollution. \citet{letelier2021solving} considered the same objective for non-recyclable bins. Similarly, other authors aimed at minimizing the number of places conditioned for installing bins in a city~\citep{adeleke2021efficient,gallardoX2015methodology,ghianiX2012capacitated,ghianiX2014impact,gilardino2017combining,hemmelmayrX2017periodic,kaoX2002shortest,kao2013spatial,maraqa2018optimization,nevrly2019municipal,nevrly2021location,ratkovic2016planning,vu2018parameter}. \citet{karadimasX2005gis} and \citet{karadimas2008gis} proposed locating a bin at every block, and then eliminating bins collecting a negligible amount of waste to diminish the number of points that should be visited by the collection vehicle.

\subsection{User accessibility criteria}

A usual criteria related to user accessibility is minimizing the average distance between users and assigned bins~\citep{herrera2018optimization,kao2010service,rossit2019bi,rossitX2017application,rossit2018municipal,rossit2020exact,toutouh2018intelligence,toutouh2020soft,tralhaoX2010multiobjective}. Other authors considered the total sum of the walking distances (for all users) instead of the average~\citep{chang1999strategic,chang2000siting,gautam2005strategic,kaoX2002shortest,kao2013spatial,nevrly2019municipal,sheriff2017integrated}. \citet{coutinhoX2012bi}, \citet{vijay2005estimation,vijay2008gis} 
applied a more specific model, considering the walking time for users to access to their assigned bin. \citet{flahautX2002locating} used the total transportation costs, which are proportional to distance. \citet{kao2013spatial} proposed models to enhance
user accessibility when collecting recyclable material, either minimizing the maximum distance that any user must walk or the percentage of citizens located within an ``aceptable'' distance from a bin. 

Some articles studied cases where minimizing the walking distance of users to assigned bins is of paramount importance. \citet{valeo1998location} considered user accessibility as critical for the success of implementing a recycling program. Two accessibility models were analyzed:
i) users are supposed to walk to the bins; thus, the model locates several bins within walking distance from households, and ii) users are supposed to access by car; thus, bins are located only on commercial parking lots. 
\citet{linX2011model} studied a 
similar scenario, where 
bins can be visited by the collection vehicle either in a day shift or in a night shift. In the problem model, each user must have at least two bins within a certain threshold distance, belonging to different shifts. Users unable to access their nearest bin before it is visited by the collection vehicle can dispose their recyclables at one of the alternative nearby bins in the other shift. \citet{carlos2021design} considered as critical the accessibility of users when installing waste bins in a city with no formal MSW collection network. To facilitate that users get used to bins, the transportation distance was minimized using two storage levels: first, users deposed their waste in bins located at their courtyard, and then, waste was transported by specialized workers to larger collection points, to be collected by vehicles. Some articles considered the distance between users and their assigned bin
as a feasibility constraint
, i.e., users cannot be assigned to a bin located beyond a certain threshold value \citep{adeleke2021efficient,blazquez2020network,cavallin2020application,coutinhoX2012bi,erfani2017novel,erfaniX2018using,ghianiX2012capacitated,ghianiX2014impact,gilardino2017combining,rossit2020exact,toutouh2020soft,tralhaoX2010multiobjective,vu2018parameter} 

Other articles have included the minimization 
the distance between a user and the assigned bin multiplied by the weight of the waste the user generates~\citep{aremu2012case,kim2013restricted,kim2015integrated,kim2015case,nevrly2021location}. A few articles directly required the users to be assigned to the nearest bin 
considering that this is the usual behavior that citizens adopt in the real-world~\citep{di2014integration,ghianiX2012capacitated,ghianiX2014impact,gilardino2017combining,kao2010service,kao2013spatial}. However, an interesting approach was proposed by~\citet{barrena2019optimizing,barrena2020solidarity}, assuming that the users have a supportive behavior regarding respecting the bin where they are assigned, even though it is not the closest one. 

Another problem variant consists in installing new bins, near collection sites of an existing network of bins
\citep{adedotun2020improving,boskovic2015fast,lopez2008optimizing,paul2017using}. The main motivation for this constraint is that users find easier to adapt to new location of bins if they are not far from the previous locations~\citep{erfani2017novel,zamorano2009planning}.

Finally, some researches have considered user accessibility under the assumption that each bin serves a certain nearby area. Optimization criteria included maximizing the geographic area~\citep{nithya2012optimal,paul2017using,vu2018parameter} or the number of users~\citep{aremu2012case,chang1999strategic,chang2000siting,erfaniX2018using,kao2010service,cubillos2020solution,letelier2021solving,vidovic2016two}. \citet{erfaniX2018using} and \citet{vidovic2016two} presented an interesting approach for estimating the population served by a bin. The models take into account the number of users of a certain bin, modeled as a decay function based on the distance to the bin and assuming that the farther the citizens must walk to the bin, the less likely they are to use it.

\subsection{Routing collection criteria}

After storage, the immediate stage of the reverse supply chain of MSW is the collection of waste. Thus, several authors have included in the optimization model of the bin location problem some criteria linked to the collection stage in order to diminish the posterior routing costs. 

Collection frequency (i.e., how often the bins are visited and emptied by the collection vehicle) is a very frequent consideration in the related literature. A low collection frequency implies that the needed storage capacity of bins (proportional to investment cost) will be also large. However, a very high collection frequency implies extra routing costs, because vehicles have to visit the bins very often~\citep{hemmelmayrX2013models,rossit2020exact}. Several authors included the trade-off between collection frequency and cost in their models. \citet{rossit2018municipal,rossit2019bi,rossit2020exact} aimed at minimizing the required collection frequency when designing the collection network of bins while also minimizing the purchasing cost of bins. \citet{blazquez2020network} solved different scenarios for different collection frequencies and found an approximately linear correlation: when the frequency decreases from every day collection to collection every three days and the cost of the collection network almost tripled. Those previous findings are in line with the results obtained by~\citet{cavallin2020application}: the cost of the network of bins increased by 50\% when modifying the collection frequency of dry waste from four times per week to three times per week. Similarly, \citet{nevrly2021location} aimed at minimizing the total collection service time, i.e., the total time (in the planning horizon) dedicated by the collection vehicle to empty the bins, which is directly associated with the collection frequency. \citet{gonzalez2002model} minimized the collection frequency as a second objective in a lexicographic approach in which the first objective was to maximize the amount of glass collected. In the context of a two-shift collection plan, \citet{linX2011model} aimed at minimizing the number of bins that have to be collected within each shift of the collection vehicle.
\blue{An interesting result was obtained by~\cite{carlos2016optimization} who was able to reduce the required collection frequency of recyclable material as a result of improving the location of bins to enhance user accessibility which lead to a reduction in the risk of overflow.}

The type of bin is also related to posterior routing costs, since decision-makers generally prefer to use the same type of bin across a certain area. The rationale behind this idea are the constraints existing in mechanized self-loading collection vehicles, e.g., limit capacity of vehicles and technical specification of the unloading mechanism~\citep{erfaniX2018using}. For example, if two bins that require different types of vehicles are located in the same site, this site has to be visited at least by two vehicles. The model must introduce a lower bound in the number of collection routes connecting the site, resulting in an increased routing cost. \citet{ghianiX2014impact} prevented bins that require different collection vehicles to be emptied to be located in the same site. Similarly, \citet{khan2016allocation} used a unique size of bins to avoid inconveniences for the collection phase. 

Another relevant aspect is the vehicle accessibility to the bins
considering that 
certain streets are not adequate for collection vehicles, either because they are extremely narrow, dirty, or the asphalt is in very poor conditions. Thus, bins should not be located in those streets~\citep{carlos2016optimization,boskovic2015fast,lopez2008optimizing,vu2018parameter}. Some models locate bins
only in main streets~\citep{aremu2016modeling,zahan2020multi} or road intersections~\citep{adedotun2020improving,khan2016allocation,vijay2008gis}. 
Other works have considered environment-related routing criteria.
\citet{aremu2012case} and \citet{aremu2012framework} proposed choosing the best location for bins taking into account the cost of fuel of the collection stage and the pollution emissions of the vehicles in the problem model and later in a sensitivity analysis to analyze the optimal number of bins to install~\citep{aremu2012case}.

A few articles directly addressed the integrated problem of locating bins and designing the collection routes simultaneously, in a unique optimization process. Thus, these works usually included some metric of the routing efficiency. For example, minimizing the idle time of vehicles (i.e., time for performing collection and travel time between waste bins)~\citep{vidovic2016two}, travel costs (per distance unit)~\citep{chang1999strategic,chang2000siting,hemmelmayrX2013models,hemmelmayrX2017periodic,jammeliX2019bi,kim2015integrated,sheriff2017integrated,vidovic2016two,yaakoubi2018heuristic} or travel distance of the vehicles~\citep{aka2018fuzzy,cubillos2020solution}, or fixed costs of the fleet of vehicles~\citep{hemmelmayrX2017periodic, kim2015integrated}. These integrated approaches provide a valuable holistic view of the problem, thus they are further described in Section~\ref{sec:integrated}.

\subsection{Other criteria}

Some works have included other particular criteria for bin location, not included in the previous
categories. 
Some models considered
that the collection network of bins may not receive the total amount of waste produced in an area, 
because it competes
with 
informal or sporadic systems, such as 
recycling campaigns by governmental or non-governmental organizations or informal workers~\citep{ferronato2020assessment,rathore2020location,toutouh2020soft,ugwuishiwu2020gis}. 
\citet{aremu2012case,aremu2012framework} and \citet{toutouh2018intelligence,toutouh2020soft} included the objective of maximizing the 
collected waste.
\citet{gonzalez2002model} 
used this objective
for the special case of glass collection, which competes with the (unclassified) formal collection system.
Instead of maximizing the recyclable material collected, 
\citet{hemmelmayrX2017periodic,vidovic2016two} proposed maximizing the profit obtained from the selling price of the collected recyclable material. \citet{sheriff2017integrated} considered that the recyclable material is collected from users by specialized agencies. The agencies deliver the material to the initial collection points and receive as a compensation an economic incentive that varies depending on the quality of the material delivered. This incentive is included as a minimization cost in the objective function of the model.

Related works also considered
criteria associated with the 
semi-obnoxiousness nature of the bins (the NIMBY phenomenon). \citet{coutinhoX2012bi} and \citet{tralhaoX2010multiobjective} 
optimized
the number of citizens within the \textit{push} and \textit{pull} distances of a bin. The push 
criterion 
minimizes
the number of users too near of the bins and the pull 
criterion 
minimized
the number of users too far from the bins. Similarly, \citet{flahautX2002locating} considered two different terms in the objective function: a term that aims at minimizing the distance between users and assigned bins and a second term that aims at minimizing the external costs, i.e., the negative environmental impact of the noise of collection vehicle for citizens that have a bin too near to their homes. In the sensitivity analysis, the authors found that when the magnitude of external costs increases, the bins are pushed too less populated areas. \citet{barrena2019optimizing} considered that the company in charge of the waste collection service should reduce the cost for citizens that are located near a bin, for the inconveniences it may cause. Also connected with the environmental impact generated by the bins, \citet{ahmed2006solid}, \citet{zahan2020multi} and \citet{ugwuishiwu2020gis} aimed at preventing the bins from being located near busy places, such as educational places and medical centers. \citet{ahmed2006solid} and \citet{ugwuishiwu2020gis} also identified water streams as an area that should be kept free from bins. In the context of locating bins in a historic neighborhood, \citet{tralhaoX2010multiobjective} avoided locating collection points near historical buildings, to not affecting the cultural heritage of the neighborhood. 

Finally, a particular study by~\citet{lopez2008optimizing} studied a case where paper and cardboard community bins tend to overflow in highly commercial areas. Using special kerbside collection for paper and cardboard was proposed in areas with many small businesses to reduce the amount of recyclable material stored in the community bins. Moreover, the bins distribution was also modified and bins were moved towards nearby residential areas with less commercial activities.

\section{Resolution method}
\label{sec:resolution_method}

This section reviews the different resolutions methods proposed in the related works to solve the waste bins location problem. First, an overall description of the analysis and categorization is presented. Then, a specific discussion is performed for each identified category.

\subsection{General discussion}

Several approaches have been applied for solving the MSW bins location problem. Some articles proposed exact methods. However, these approaches may not be able to solve large instances since waste bins location problem is a NP-hard problem. Thus, non-exacts methods (i.e., heuristic and metaheuristic approaches~\cite{Nesmachnow2014}) 
have also been applied to solve the problem. A common practice is to apply exact approaches for solving small scenarios and validating the heuristics/metaheuristics and, after that, applying the non-exact methods to solve large scenarios. Since a large part of the input data of the applied models consists in geo-referenced information, the capabilities of ArcGIS have been used, not only for processing the information and analyzing the results, but also for solving the optimization problem. GIS technologies support many resolution algorithms for facility location models, from exact $p$-medians models to heuristic approaches. Finally, a few works used other specific approaches, e.g., solving the problem by a manual approach based on some processed information (based on GIS or algebraic formulae), applying multicriteria methods such as AHP, or applying non-formally documented resolution approaches.  

Another relevant aspect when solving real-world scenarios is considering uncertainty in the input parameters. Several articles have considered variations in the input parameters or, at least, have evaluated different typical values of waste generation rate, maximum threshold distance that users are willing to carry their waste to a bin, or  maximum number of bins to install. Thus, a specific subsection is included for discussing works addressing uncertainty.

This review analyzes the resolution approaches used in the related works discussing the following topics: \textit{exact approaches}, for works that applied exact resolution methods; \textit{heuristic and metaheuristic approaches}, for works that applied non-exact methods, the \textit{GIS-based and other approaches}, for works that used GIS-based and other location methods; and \textit{approaches considering uncertainty}, for works that consider uncertainty in some input parameter.

\subsection{Exact approaches}

Many articles applied 
MILP
models for solving the waste bins location problem. More than a half used single-objective models. Other articles applied multi-objective models to combine different criteria, e.g., a weighting sum approach~\citep{cubillos2020solution,herrera2018optimization,nevrly2019municipal,nevrly2021location,rossit2018municipal,rossit2019bi,rossit2020exact,tralhaoX2010multiobjective}, $\varepsilon$-constraint \citep{coutinhoX2012bi,nevrly2021location,rossitX2017application,rossit2020exact}, lexicographic optimization~\citep{gonzalez2002model,rossit2020exact} or goal programming~\citep{aka2018fuzzy,tralhaoX2010multiobjective}.
In several works, MILP exact models were used for solving small instances to validate heuristic approaches, later used to address larger scenarios~\citep{cubillos2020solution,ghianiX2012capacitated,ghianiX2014impact,hemmelmayrX2013models,hemmelmayrX2017periodic,kim2013restricted,vidovic2016two}. 

A few works presented resolution approaches that combine linear programming models with other optimization tools. \citet{aka2018fuzzy} combined MILP models with fuzzy goals for the optimization criteria. \citet{jammeliX2019bi} used an stochastic MILP model embedded in a heuristic approach to solve an integrated problem that combines defining the bins location and the routing collection schedules. Also solving an integrated problem, \citet{vidovic2016two} implemented linear programming model as part of a two-phase heuristic procedure. \citet{rathore2019allocation} and \citet{rathore2020location} applied a single-objective MILP approach only to determine the number of required bins in each region and, then, the location inside each region is determined whit ArcGIS.

\subsection{Heuristic and metaheuristic approaches}

Diverse heuristic/metaheuristic approaches have been proposed 
to solve the waste bins location problem. \citet{bautista2006modeling} proposed 
an evolutionary algorithm (EA) 
and a GRASP heuristic for efficiently solving real instances of the problem. \citet{toutouh2018intelligence} applied the PageRank voting algorithm and 
NSGA-II
to solve a multiobjective version of the problem. Later, other multiobjective approaches were applied: a multi-objective PageRank algorithm and 
SPEA-2
\citep{toutouh2020soft}.
\citet{barrena2019optimizing} proposed a greedy algorithm, later improved by including a
local search to the computed solution~\citep{barrena2020solidarity}. \citet{adeleke2021efficient} applied a heuristic based on a Lagrangian relaxation of a linear programming model to solve the problem. \citet{vijay2005estimation} programmed a greedy heuristic in Arc Macro Language of ArcGIS that uses triangulated irregular networks and assigns users to the nearest bin considering path slopes.

Multi-stage heuristics have been proposed
to solve different parts of the problem sequentially. In the 
constructive heuristic by~\citet{ghianiX2012capacitated,ghianiX2014impact}, the first stage determines the sites for bins from a set of potential locations and 
assigns
users to sites, and the second stage allocates bins to selected places. 
\citet{di2014integration} proposed a similar 
heuristic: first, deciding the places to locate the bins and, second, determining the size of the bins 
for
each place. \citet{kim2013restricted} proposed two multi-stage heuristics for solving large instances of the problem: i) a multi-stage branch and bound and ii) a drop heuristic. In both methods, first, the problem is solved considering static user-bin assignments and then, the solution is improved by allowing dynamic assignments, i.e., the users can be assigned to different bins on different days within the time horizon. 

Several articles applied metaheuristics to 
the integrated problem of locating waste bins and designing collection routes, 
including
EAs~\citep{chang2000siting}, memetic algorithms 
\citep{yaakoubi2018heuristic}, Variable Neighborhood Search (VNS)~\citep{cubillos2020solution,hemmelmayrX2013models}, Adaptive Large Neighborhood Search~\citep{hemmelmayrX2017periodic}, 
and 
clustering
\citep{jammeliX2019bi}. 
These approaches are 
described in Section~\ref{sec:integrated}. 

\subsection{GIS-based and other approaches}

In the last decades, GIS software evolved from products designed to organize and show geographic information to complex integrated platforms for 
acquiring, storing, analyzing, and visualizing geospatial data
\citep{li2020real}. 
Curent GIS software includes
solving optimization problems using predefined tools, or building specific solvers using integrated programming languages.

Several authors have used ArcGIS Network Analyst, an application integrated in ArcGIS to efficiently analyze and processed realistic network models~\citep{zamorano2009planning}. \citet{aremu2016modeling} used the Location-Allocation tool in ArcGIS Network Analyst to locate waste bins and assign users to the bins considering typical footpaths of the citizens in the area of study. \citet{erfani2017novel} applied the Minimize Facilities model of ArcGIS Network Analyst to determine the number of bins that cover maximum population
within a 
maximum walking distance, from a potential set of GAPs. The approach was enhanced by applying the Maximized Capacitated Coverage model to assess the population and the waste received by the facilities~\citep{erfaniX2018using}. \citet{vu2018parameter} applied the Maximize Coverage and Minimize Facility models to compute solutions that considered the trade-off between costs and QoS provided to the citizens. Some works developed 
applications for optimization in GIS embedded languages. \citet{vijay2005estimation} developed a greedy heuristic in Arc Macro Language of ArcGIS. \citet{gautam2005strategic} and \citet{vijay2008gis} programmed a $p$-median model in the same programming language. \citet{aremu2012case} and \citet{aremu2012framework} programmed a $p$-median model in TransCAD software~\citep{lu2008intro}. 

Another set of works 
used GIS
for processing information 
to analyze
an existing network of bins and suggest modifications to improve the coverage by a manual trial-and-error method. \citet{boskovic2015fast} and \citet{lopez2008optimizing,lopez2009containerisation} suggested some modifications after carefully considering the waste generation by citizens, stores, and institutions for different seasons of the year. The analysis by \citet{zamorano2009planning} concluded that there was a lack of bins for the recyclable network and an excessive number of bins for the non-recyclable network.
After a survey to carefully estimate waste generation, \citet{khan2016allocation} relocated the bins in the studied area to reduce the posterior collection costs. \citet{ferronato2020assessment} analyzed the MSW system considering the contribution of informal waste pickers and \citet{karadimas2008gis} presented a regression model to accurately estimate the waste generation rate and the required number of bins.

Finally, two works used the multicriteria Analytical Hierarchy Procedure (AHP). \citet{aremu2012framework} used AHP to select the best solution based on economic, social, and environmental criteria among several outcomes of the $p$-median models with different number of bins. \citet{zahan2020multi} applied AHP in a pilot case to choose among three different locations to place the bin. 

\subsection{Approaches considering uncertainty}

Another important aspect in the resolution process of the waste bins location problem is the consideration of uncertainty in the input parameters. 

One of the main sources of uncertainty 
is waste generation rate, which is not easy to estimate since it depends on population density, commercial activities in the area~\citep{karadimas2006municipal,karadimas2008gis}, and it is affected by seasonal variations~\citep{boskovic2015fast,carlos2021design}.
\citet{jammeliX2019bi} considered a stochastic normally distributed number of citizens within the influence area of each collection point and stochastic waste generation,
integrated in the 
resolution process via a transformed MILP formulation. \citet{kim2013restricted,kim2015case,kim2015integrated}
considered the variation of waste generation for different days within the planning time horizon, 
allowing
the model to assign users to different bins on each day. 
Another approach considers 
deterministic generation, but sensitivity analysis are performed by solving the waste bins location problem for different common values of waste generation rate~\citep{letelier2021solving,rathore2020location,rossit2018municipal,rossit2019bi,toutouh2018intelligence,toutouh2020soft}. Finally, other articles considered a simpler strategy using a safety factor in the usage of bins, i.e., an specified proportion of their capacity must remain unused to absorb potential increments in the waste generation rate~\citep{boskovic2015fast,ferronato2020assessment,rathore2020location}. However, the safety factor strategy must be carefully planned. \citet{perez2017methodology} found that unused overcapacity increases the environmental impact of the waste bins network. \citet{aremu2012framework} and \citet{erfaniX2018using} considered the unused 
capacity as a negative factor 
that increments
the purchasing and maintenance costs of bins, an adverse effect on aesthetic, and decreases the efficiency of the collection.
%
\citet{flahautX2002locating} found that estimating the waste generation rate may not lead to better results.
Similar quality of service was computed when using two different estimations of the generation rate:
a simple approximation based on population density and a more accurate estimation considering additional features of users (users per household, socioeconomic level, and age ranges), gathered in an specific survey. Thus, researchers and authorities must carefully consider whether is worth spending time and effort in collecting additional information from users or not.

Another important parameter considered in sensitivity analysis is the maximum tolerable distance between users and their assigned bins. Related works have considered a wide range of distance threshold values, from short distances (75\,m)~\citep{zamorano2009planning} to medium distances (up to 500\,m)~\citep{di2014integrationpilotcase,parrotX2009municipal}. The maximum distance that citizens are willing to transport their waste is highly site-specific~\citep{boskovic2015fast,vu2018parameter} and it depends on the type of collection system (kerbside, drop-off stations, etc.)~\citep{gallardoX2015methodology}. Several articles reported sensitivity analysis to evaluate models
for different common distance thresholds~\citep{adeleke2021efficient,aremu2012case,aremu2016modeling,cavallin2020application,kaoX2002shortest,erfani2017novel,di2014integrationpilotcase,ghianiX2012capacitated,ghianiX2014impact,letelier2021solving,linX2011model,oliaeibreakdown,ratkovic2016planning,vu2018parameter}.

Other articles limited the number of 
bins
to install in the studied scenario, according to
sensitivity analysis~\citep{cubillos2020solution,hemmelmayrX2013models,kao2013spatial,kim2013restricted,kim2015case,kim2015integrated,vu2018parameter}. \citet{coutinhoX2012bi} bounded the installment costs (purchase of bins and installation of bins requiring collection vehicle adaptation).
The model was solved for different threshold values for the 
cost, 
to analyze the its impact on the quality of 
solutions.

A few articles have performed sensitivity analysis in less common parameters, such as the fixed collection frequency 
accumulated waste
\citep{carlos2021design,cavallin2020application,hemmelmayrX2013models,rathore2020location}, the number of periods in the planning horizon~\citep{kim2013restricted,kim2015integrated}, the bins storage capacity~\citep{adeleke2021efficient,hemmelmayrX2013models}, the purchasing cost of bins~\citep{hemmelmayrX2013models}, the 
frequency of maintenance tasks on bins~\citep{gilardino2017combining}, the maximum distance between collection points~\citep{blazquez2020network}, or the nuisance of the noise generated by collection vehicles~\citep{flahautX2002locating}.

Finally, two articles considered fuzzy goals in integrated approaches. \citet{chang2000siting} considered three objectives (population served, average walking distance between users and assigned bins, and routing distance of collection vehicles) as fuzzy planning goals to compute a set of compromising solutions. \citet{aka2018fuzzy} also used fuzzy objective functions to maximize the number of citizens served and minimize the routing distance of the collection vehicles.

\section{Integrated approaches with other stages of the reverse MSW supply chain}
\label{sec:integrated}

\blue{The traditional approach for decision-making in the MSW system solved problems on a sequential or hierarchical fashion, i.e, first 
locating the bins and then optimize the routes needed for waste collection. However, the main goal of designing a useful MSW system providing both cost savings and proper QoS can be achieved by applying integrated models to solve both problems simultaneously, in the same optimization process~\citep{kim2015integrated}. The resolution is computationally challenging, since it involves addressing two NP-hard problems simultaneously.}

The advantages of 
integrated approaches
are noteworthy,
considering
the 
interdependence
between bins network design and waste collection routes. 
The geographical distribution of bins clearly affects the routing schedule, since the selected locations must be visited by the collection vehicles \citep{cubillos2020solution}. The storage capacity of the installed bins also influences the 
collection cost,
determining 
the required 
frequency to empty the bins and avoid overflowing~\citep{hemmelmayrX2013models}. Moreover, the routing cost is affected by the type of bins used, i.e., special bins may require compatible vehicles to be emptied~\citep{ghianiX2014impact}. Thus, unlike traditional sequential methods, the holistic perspective of integrated approaches captures the connection between the network design and the collection route 
in the decision-making process, reducing the overall cost of the system~\citep{bingX2016research}. 
This section describes integrated approaches 
that 
provides the more comprehensive, challenging, and interesting solutions for MSW systems.

Articles considering integrated approaches are relatively scarce. 
\blue{The first integrated approach was proposed by~\citet{chang1999strategic} for location-routing of recyclable materials. 
A multiobjective linear programming model was proposed, }
considering three objective functions: maximization of population served, and minimization of total walking distance from household to recycling drop-off stations and the total driving distance of the collection vehicles. A simple EA was proposed to solve the problem, later extended to consider a fuzzy fitness function~\citep{chang2000siting}.
The integrated approach by~\citet{hemmelmayrX2013models} considered waste generation concentrated in the places where bins are to be installed. 
Bins location, collection frequency, and routing schedule were solved jointly using a MILP model for small instances and VNS for larger instances. The VNS solved the waste collection routing problem by assigning collection frequencies to each collection site, and re-optimizing the bins location to obtain a new minimal cost  
\blue{by applying a local search heuristic}.
For small instances, the VNS 
was compared with a MILP model, obtaining accurate results. For larger instances, the integrated VNS 
\blue{outperformed a sequential approach solving the bin location and then the waste collection, or viceversa}.

\citet{kim2015integrated} considered a dynamic allocation of users to waste bins
in different days. The integrated procedure applied Tabu Search (TS) for bins location. Each time the location of bins or the users-bins allocation change, the routing cost is calculated applying another TS 
that searches 
on neighborhoods built using 2-opt/3-opt heuristics. The integrated approach outperformed hierarchical methods applying TS and a cluster-first route-second heuristic with a local search improvement step for setting the routing schedule. \citet{sheriff2017integrated} also considered three stages for plastic collection: location of waste bins, location of transfer plants, and the routing schedules from waste bins to transfer plants. The integrated approach improved over solutions computed by a three-stages sequential approach. Flexibility either in the potential location of vehicle depots or the visit schedules of the collection points reduced the total costs, without increasing the number of depot locations.

\citet{hemmelmayrX2017periodic} presented a location-routing approach 
for multi-agency
recycling campaigns.
The optimization model aimed at simultaneously define which agency should own a recyclable bin, the capacity of the bins to be installed in each place, and set the weekly schedule and routes for the collection vehicles. 
A MILP formulation was proposed for small instances and Adaptive Large Neighborhood Search (ALNS) for larger instances. 
ALNS obtained near optimal solutions when compared with the MILP formulation. A sensitivity analysis for the available vehicle capacities, the visiting schedules, or the capacity of installed bins was also presented.

\citet{aka2018fuzzy} presented an integrated approach aimed at maximizing the total number of dwellings served by a network of recyclable bins and minimize the total travel distance by the collection vehicle. A fuzzy goal programming approach was proposed for solving a MILP formulation of the problem using goals previously estimated by experts. The allocation of users to each bin is considered to be given beforehand and it is not part of the optimization process. 

\citet{jammeliX2019bi} simultaneously allocated bins to predefined locations and defined collection routes, considering stochastic waste generation. The resolution approach applied two stages: a $k$-means clustering to group bins into sectors, and an exact model to determine both the number of bins and the collection route for each sector, assuming a daily collection frequency. The stochastic waste generation rate influenced the required capacity of bins to locate in each place, the usage of the vehicle capacity along the route, and the collection time that takes to the vehicle to empty a bin. Uncertainty was coped by several transformations for solving the model in the second stage.

The model by \citet{cubillos2020solution} considered each household as a potential site for locating a bin for recyclable material. The goal was finding a network of bins to simultaneously maximize the number of covered households, i.e., located within a certain threshold distance from a bin, and minimize the collection route, which was approximated solving a Traveling Salesman Problem without considering the capacity or other real-world constraints. This simplification allowed obtaining fast estimations for the collection cost without solving a time-consuming inventory-routing problem. For small instances, the VNS was compared with a MILP model obtaining accurate results. Then, the VNS was used to solve larger instances and a real-world case study.

Finally. other two works presented MILP models for the integrated problem that were only applied to small toy instances~\citep{vidovic2016two,yaakoubi2018heuristic}. Then, these two works, proposed heuristics for solving both problems in a sequential fashion for realistic instances.

\section{Conclusions}
\label{sec:conclusion}
Waste management is 
critical
for modern cities, since it has direct implications on the environmental, social, and economic welfare of citizens. Among the diverse stages of the reverse supply chain of MSW, this article addressed the waste bins location problem,
consisting
in finding the best locations for community bins in an urban area to store the waste before they are collected by the collection vehicle. This problem has an important impact on the overall efficiency of the logistic chain, since it is the entry point to the MSW system. Thus, a poor planning on this stage can diminish the amount of waste received by the MSW system of a city. Moreover, the distribution and the capacity of bins influence other stages of the reverse supply chain, e.g., the design and frequency of the collection routes and the required capacity of intermediate and processing facilities. There is also evidence that a correct arrangement of bins can encourage the community to correctly classify the waste at source, which usually is closely associated with the success of recycling programs.

The review performed a thorough analysis of 76 related works, carefully selected from more than two hundred articles retrieved from 
well-known 
scientific databases.
After presenting a general description of the reverse supply chain of MSW and the importance of the waste bin location problem for the MSW system, the selected articles were \blue{studied through bibliometric analysis, considering the addressed scenarios, pursued criteria, resolution approaches, and issues related to the waste collection stage. In turn, articles were}
reviewed taking into account: i) the optimization criteria used, ii) the resolution approach, iii) the consideration of uncertainty, iv) the application of integrated approaches to address bin location and collection routing simultaneously.

The review of related works has suggested some interesting topics to be addressed in future researches. An important line for future work is incorporating uncertainty in the conceptual models and the resolution approaches. The main factor subject to uncertainty is the waste generation, mainly due to seasonal and citizens consumption pattern variations. Although a few works have performed sensitivity analysis based on typical values of this parameter, only one article considered waste generation as stochastic and, thus, applied an stochastic approach. Another promising research line 
is to continue developing integrated approaches considering the trade-off between the storage and the waste collection stages of the MSW logistic chain. Most articles in the literature assumed a sequential approach of both stages, although there is evidence that integrated approaches are able to outperform sequential approaches due to a better characterization of the interrelation between the underlying optimization problems.

\bibliographystyle{cas-model2-names}
\bibliography{mybibfile_final.bib}

\begin{thebibliography}{116}
\expandafter\ifx\csname natexlab\endcsname\relax\def\natexlab#1{#1}\fi
\providecommand{\url}[1]{\texttt{#1}}
\providecommand{\href}[2]{#2}
\providecommand{\path}[1]{#1}
\providecommand{\DOIprefix}{doi:}
\providecommand{\ArXivprefix}{arXiv:}
\providecommand{\URLprefix}{URL: }
\providecommand{\Pubmedprefix}{pmid:}
\providecommand{\doi}[1]{\href{http://dx.doi.org/#1}{\path{#1}}}
\providecommand{\Pubmed}[1]{\href{pmid:#1}{\path{#1}}}
\providecommand{\bibinfo}[2]{#2}
\ifx\xfnm\relax \def\xfnm[#1]{\unskip,\space#1}\fi
\bibitem[{Adedotun et~al.(2020)Adedotun, Sridhar and
  Coker}]{adedotun2020improving}
\bibinfo{author}{Adedotun, A.}, \bibinfo{author}{Sridhar, M.},
  \bibinfo{author}{Coker, A.}, \bibinfo{year}{2020}.
\newblock \bibinfo{title}{Improving municipal solid waste collection system
  through a {GIS} based mapping of location specific waste bins in {I}badan
  {M}etropolis, {N}igeria}.
\newblock \bibinfo{journal}{Journal of Solid Waste Technology \& Management}
  \bibinfo{volume}{46}, \bibinfo{pages}{360–371}.
\newblock \DOIprefix\doi{10.5276/JSWTM/2020.360}.
\bibitem[{Adeleke and Ali(2021)}]{adeleke2021efficient}
\bibinfo{author}{Adeleke, O.}, \bibinfo{author}{Ali, M.}, \bibinfo{year}{2021}.
\newblock \bibinfo{title}{An efficient model for locating solid waste
  collection sites in urban residential areas}.
\newblock \bibinfo{journal}{International Journal of Production Research}
  \bibinfo{volume}{59}, \bibinfo{pages}{798--812}.
\newblock \DOIprefix\doi{10.1080/00207543.2019.1709670}.
\bibitem[{Ahmed et~al.(2006)Ahmed, Muhammad and Sivertun}]{ahmed2006solid}
\bibinfo{author}{Ahmed, S.}, \bibinfo{author}{Muhammad, H.},
  \bibinfo{author}{Sivertun, A.}, \bibinfo{year}{2006}.
\newblock \bibinfo{title}{Solid waste management planning using {GIS} and
  remote sensing technologies case study {Aurangabad City, India}}, in:
  \bibinfo{booktitle}{2006 International Conference on Advances in Space
  Technologies}, pp. \bibinfo{pages}{196--200}.
\newblock \DOIprefix\doi{10.1109/ICAST.2006.313826}.
\bibitem[{Aka and Aky{\"u}z(2018)}]{aka2018fuzzy}
\bibinfo{author}{Aka, S.}, \bibinfo{author}{Aky{\"u}z, G.},
  \bibinfo{year}{2018}.
\newblock \bibinfo{title}{Fuzzy goal programming approach on location-routing
  model for waste containers}.
\newblock \bibinfo{journal}{International Journal of Industrial and Systems
  Engineering} \bibinfo{volume}{29}, \bibinfo{pages}{413--427}.
\newblock \DOIprefix\doi{10.1504/IJISE.2018.094265}.
\bibitem[{Ali and Ahmad(2019)}]{ali2019spatial}
\bibinfo{author}{Ali, S.}, \bibinfo{author}{Ahmad, A.}, \bibinfo{year}{2019}.
\newblock \bibinfo{title}{Spatial susceptibility analysis of vector-borne
  diseases in {KMC} using geospatial technique and {MCDM} approach}.
\newblock \bibinfo{journal}{Modeling Earth Systems and Environment}
  \bibinfo{volume}{5}, \bibinfo{pages}{1135--1159}.
\newblock \DOIprefix\doi{10.1007/s40808-019-00586-y}.
\bibitem[{Aremu and Sule(2012)}]{aremu2012case}
\bibinfo{author}{Aremu, A.}, \bibinfo{author}{Sule, B.}, \bibinfo{year}{2012}.
\newblock \bibinfo{title}{A case study evaluation of the impacts of optimised
  waste bin locations in a developing city}.
\newblock \bibinfo{journal}{Civil Engineering and Environmental Systems}
  \bibinfo{volume}{29}, \bibinfo{pages}{137--146}.
\newblock \DOIprefix\doi{10.1080/10286608.2012.672411}.
\bibitem[{Aremu et~al.(2012)Aremu, Sule, Downs and
  Mihelcic}]{aremu2012framework}
\bibinfo{author}{Aremu, A.}, \bibinfo{author}{Sule, B.},
  \bibinfo{author}{Downs, J.}, \bibinfo{author}{Mihelcic, J.},
  \bibinfo{year}{2012}.
\newblock \bibinfo{title}{Framework to determine the optimal spatial location
  and number of municipal solid waste bins in a developing world urban
  neighborhood}.
\newblock \bibinfo{journal}{Journal of Environmental Engineering}
  \bibinfo{volume}{138}, \bibinfo{pages}{645--653}.
\newblock \DOIprefix\doi{10.1061/(ASCE)EE.1943-7870.0000513}.
\bibitem[{Aremu and Vijay(2016)}]{aremu2016modeling}
\bibinfo{author}{Aremu, A.}, \bibinfo{author}{Vijay, R.}, \bibinfo{year}{2016}.
\newblock \bibinfo{title}{Modeling indigenous footpath and proximity cut-off
  values for municipal solid waste management: a case study of ilorin,
  nigeria}.
\newblock \bibinfo{journal}{Procedia Environmental Sciences}
  \bibinfo{volume}{35}, \bibinfo{pages}{51--56}.
\newblock \DOIprefix\doi{10.1016/j.proenv.2016.07.005}.
\bibitem[{Barrena et~al.(2019)Barrena, Canca, Ortega and Piedra de~la
  Cuadra}]{barrena2019optimizing}
\bibinfo{author}{Barrena, E.}, \bibinfo{author}{Canca, D.},
  \bibinfo{author}{Ortega, F.}, \bibinfo{author}{Piedra de~la Cuadra, R.},
  \bibinfo{year}{2019}.
\newblock \bibinfo{title}{Optimizing container location for selective
  collection of urban solid waste}, in: \bibinfo{booktitle}{9th International
  Conference on Waste Management and the Environment}, pp.
  \bibinfo{pages}{1--9}.
\newblock \DOIprefix\doi{10.2495/WM180011}.
\bibitem[{Barrena et~al.(2020)Barrena, Canca, Ortega and Piedra-de-la
  Cuadra}]{barrena2020solidarity}
\bibinfo{author}{Barrena, E.}, \bibinfo{author}{Canca, D.},
  \bibinfo{author}{Ortega, F.}, \bibinfo{author}{Piedra-de-la Cuadra, R.},
  \bibinfo{year}{2020}.
\newblock \bibinfo{title}{Solidarity behavior for optimizing the waste
  selective collection}.
\newblock \bibinfo{journal}{International Journal of Sustainable Development
  and Planning} \bibinfo{volume}{15}, \bibinfo{pages}{133--140}.
\newblock \DOIprefix\doi{10.18280/ijsdp.150202}.
\bibitem[{Bautista and Pereira(2006)}]{bautista2006modeling}
\bibinfo{author}{Bautista, J.}, \bibinfo{author}{Pereira, J.},
  \bibinfo{year}{2006}.
\newblock \bibinfo{title}{Modeling the problem of locating collection areas for
  urban waste management. an application to the metropolitan area of
  {Barcelona}}.
\newblock \bibinfo{journal}{Omega} \bibinfo{volume}{34},
  \bibinfo{pages}{617--629}.
\newblock \DOIprefix\doi{10.1016/j.omega.2005.01.013}.
\bibitem[{Bennekrouf et~al.(2020)Bennekrouf, Aggoune, Benladghem and
  Cherif}]{bennekrouf2020strategic}
\bibinfo{author}{Bennekrouf, M.}, \bibinfo{author}{Aggoune, W.},
  \bibinfo{author}{Benladghem, K.}, \bibinfo{author}{Cherif, H.},
  \bibinfo{year}{2020}.
\newblock \bibinfo{title}{A strategic approach for the optimal location of
  recycling bins in the city of {Boudjlida in Algeria}}, in:
  \bibinfo{booktitle}{2020 IEEE 13th International Colloquium of Logistics and
  Supply Chain Management (LOGISTIQUA)}, pp. \bibinfo{pages}{1--6}.
\newblock \DOIprefix\doi{10.1109/LOGISTIQUA49782.2020.9353893}.
\bibitem[{Bing et~al.(2016)Bing, Bloemhof, Ramos, Barbosa, Wong and van~der
  Vorst}]{bingX2016research}
\bibinfo{author}{Bing, X.}, \bibinfo{author}{Bloemhof, J.},
  \bibinfo{author}{Ramos, T.}, \bibinfo{author}{Barbosa, A.},
  \bibinfo{author}{Wong, C.}, \bibinfo{author}{van~der Vorst, J.},
  \bibinfo{year}{2016}.
\newblock \bibinfo{title}{Research challenges in municipal solid waste
  logistics management}.
\newblock \bibinfo{journal}{Waste Management} \bibinfo{volume}{48},
  \bibinfo{pages}{584--592}.
\newblock \DOIprefix\doi{10.1016/j.wasman.2015.11.025}.
\bibitem[{Blazquez and Paredes(2020)}]{blazquez2020network}
\bibinfo{author}{Blazquez, C.}, \bibinfo{author}{Paredes, G.},
  \bibinfo{year}{2020}.
\newblock \bibinfo{title}{{Network design of a household waste collection
  system: A case study of the commune of Renca in Santiago, Chile}}.
\newblock \bibinfo{journal}{Waste Management} \bibinfo{volume}{116},
  \bibinfo{pages}{179--189}.
\newblock \DOIprefix\doi{10.1016/j.wasman.2020.07.027}.
\bibitem[{Bonomo et~al.(2012)Bonomo, Dur{\'a}n, Larumbe and
  Marenco}]{bonomoX2012method}
\bibinfo{author}{Bonomo, F.}, \bibinfo{author}{Dur{\'a}n, G.},
  \bibinfo{author}{Larumbe, F.}, \bibinfo{author}{Marenco, J.},
  \bibinfo{year}{2012}.
\newblock \bibinfo{title}{A method for optimizing waste collection using
  mathematical programming: a {Buenos Aires} case study}.
\newblock \bibinfo{journal}{Waste Management \& Research} \bibinfo{volume}{30},
  \bibinfo{pages}{311--324}.
\newblock \DOIprefix\doi{10.1177/0734242X11402870}.
\bibitem[{Boskovic and Jovicic(2015)}]{boskovic2015fast}
\bibinfo{author}{Boskovic, G.}, \bibinfo{author}{Jovicic, N.},
  \bibinfo{year}{2015}.
\newblock \bibinfo{title}{Fast methodology to design the optimal collection
  point locations and number of waste bins: A case study}.
\newblock \bibinfo{journal}{Waste Management \& Research} \bibinfo{volume}{33},
  \bibinfo{pages}{1094--1102}.
\newblock \DOIprefix\doi{10.1177/0734242X15607426}.
\bibitem[{Boskovic et~al.(2016)Boskovic, Jovicic, Jovanovic and
  Simovic}]{boskovic2016calculating}
\bibinfo{author}{Boskovic, G.}, \bibinfo{author}{Jovicic, N.},
  \bibinfo{author}{Jovanovic, S.}, \bibinfo{author}{Simovic, V.},
  \bibinfo{year}{2016}.
\newblock \bibinfo{title}{Calculating the costs of waste collection: A
  methodological proposal}.
\newblock \bibinfo{journal}{Waste management \& research} \bibinfo{volume}{34},
  \bibinfo{pages}{775--783}.
\newblock \DOIprefix\doi{10.1177/0734242X16654980}.
\bibitem[{Carlos et~al.(2021)Carlos, Gallardo, Colomer and
  Barreda}]{carlos2021design}
\bibinfo{author}{Carlos, M.}, \bibinfo{author}{Gallardo, A.},
  \bibinfo{author}{Colomer, F.}, \bibinfo{author}{Barreda, E.},
  \bibinfo{year}{2021}.
\newblock \bibinfo{title}{Design of a municipal solid waste collection system
  in situations with a lack of resources: {Nikki (Benin), a Case in Africa}}.
\newblock \bibinfo{journal}{Sustainability} \bibinfo{volume}{13},
  \bibinfo{pages}{1785}.
\newblock \DOIprefix\doi{10.3390/su13041785}.
\bibitem[{Carlos et~al.(2016)Carlos, Gallardo, Peris and
  Colomer}]{carlos2016optimization}
\bibinfo{author}{Carlos, M.}, \bibinfo{author}{Gallardo, A.},
  \bibinfo{author}{Peris, M.}, \bibinfo{author}{Colomer, F.},
  \bibinfo{year}{2016}.
\newblock \bibinfo{title}{Optimization of the location of the municipal solid
  waste bins using geographic information systems}, in:
  \bibinfo{booktitle}{18th International Congress on Project Management and
  Engineering}, pp. \bibinfo{pages}{171--184}.
\newblock \DOIprefix\doi{10.1007/978-3-319-26459-2_13}.
\bibitem[{Cavallin et~al.(2020)Cavallin, Rossit, Herr{\'a}n~Symonds, Rossit and
  Frutos}]{cavallin2020application}
\bibinfo{author}{Cavallin, A.}, \bibinfo{author}{Rossit, D.},
  \bibinfo{author}{Herr{\'a}n~Symonds, V.}, \bibinfo{author}{Rossit, D.},
  \bibinfo{author}{Frutos, M.}, \bibinfo{year}{2020}.
\newblock \bibinfo{title}{Application of a methodology to design a municipal
  waste pre-collection network in real scenarios}.
\newblock \bibinfo{journal}{Waste Management \& Research} \bibinfo{volume}{38},
  \bibinfo{pages}{117--129}.
\newblock \DOIprefix\doi{10.1177/0734242X19894630}.
\bibitem[{Chang and Wei(1999)}]{chang1999strategic}
\bibinfo{author}{Chang, N.}, \bibinfo{author}{Wei, Y.}, \bibinfo{year}{1999}.
\newblock \bibinfo{title}{Strategic planning of recycling drop-off stations and
  collection network by multiobjective programming}.
\newblock \bibinfo{journal}{Environmental Management} \bibinfo{volume}{24},
  \bibinfo{pages}{247--263}.
\newblock \DOIprefix\doi{10.1007/s002679900230}.
\bibitem[{Chang and Wei(2000)}]{chang2000siting}
\bibinfo{author}{Chang, N.}, \bibinfo{author}{Wei, Y.}, \bibinfo{year}{2000}.
\newblock \bibinfo{title}{Siting recycling drop-off stations in urban area by
  genetic algorithm-based fuzzy multiobjective nonlinear integer programming
  modeling}.
\newblock \bibinfo{journal}{Fuzzy Sets and Systems} \bibinfo{volume}{114},
  \bibinfo{pages}{133--149}.
\newblock \DOIprefix\doi{10.1016/S0165-0114(98)00192-4}.
\bibitem[{Chatzouridis and Komilis(2012)}]{chatzouridis2012methodology}
\bibinfo{author}{Chatzouridis, C.}, \bibinfo{author}{Komilis, D.},
  \bibinfo{year}{2012}.
\newblock \bibinfo{title}{A methodology to optimally site and design municipal
  solid waste transfer stations using binary programming}.
\newblock \bibinfo{journal}{Resources, Conservation and Recycling}
  \bibinfo{volume}{60}, \bibinfo{pages}{89--98}.
\newblock \DOIprefix\doi{10.1016/j.resconrec.2011.12.004}.
\bibitem[{Cornu{\'e}jols et~al.(1991)Cornu{\'e}jols, Sridharan and
  Thizy}]{cornuejolsX1991comparison}
\bibinfo{author}{Cornu{\'e}jols, G.}, \bibinfo{author}{Sridharan, R.},
  \bibinfo{author}{Thizy, J.}, \bibinfo{year}{1991}.
\newblock \bibinfo{title}{{A comparison of heuristics and relaxations for the
  Capacitated Plant Location Problem}}.
\newblock \bibinfo{journal}{European Journal of Operational Research}
  \bibinfo{volume}{50}, \bibinfo{pages}{280--297}.
\newblock \DOIprefix\doi{10.1016/0377-2217(91)90261-S}.
\bibitem[{Coutinho et~al.(2012)Coutinho, Tralh{\~a}o and
  Al{\c{c}}ada}]{coutinhoX2012bi}
\bibinfo{author}{Coutinho, J.}, \bibinfo{author}{Tralh{\~a}o, L.},
  \bibinfo{author}{Al{\c{c}}ada, L.}, \bibinfo{year}{2012}.
\newblock \bibinfo{title}{A bi-objective modeling approach applied to an urban
  semi-desirable facility location problem}.
\newblock \bibinfo{journal}{European Journal of Operational Research}
  \bibinfo{volume}{223}, \bibinfo{pages}{203--213}.
\newblock \DOIprefix\doi{10.1016/j.ejor.2012.05.037}.
\bibitem[{Cubillos and W{\o}hlk(2020)}]{cubillos2020solution}
\bibinfo{author}{Cubillos, M.}, \bibinfo{author}{W{\o}hlk, S.},
  \bibinfo{year}{2020}.
\newblock \bibinfo{title}{Solution of the maximal covering tour problem for
  locating recycling drop-off stations}.
\newblock \bibinfo{journal}{Journal of the Operational Research Society} ,
  \bibinfo{pages}{1--16}\DOIprefix\doi{10.1080/01605682.2020.1746701}.
\bibitem[{Dahl{\'e}n and Lagerkvist(2010)}]{dahlen2010evaluation}
\bibinfo{author}{Dahl{\'e}n, L.}, \bibinfo{author}{Lagerkvist, A.},
  \bibinfo{year}{2010}.
\newblock \bibinfo{title}{Evaluation of recycling programmes in household waste
  collection systems}.
\newblock \bibinfo{journal}{Waste Management \& Research} \bibinfo{volume}{28},
  \bibinfo{pages}{577--586}.
\newblock \DOIprefix\doi{10.1177/0734242X09341193}.
\bibitem[{Das et~al.(2019)Das, Lee, Kumar, Kim, Lee and
  Bhattacharya}]{das2019solid}
\bibinfo{author}{Das, S.}, \bibinfo{author}{Lee, S.}, \bibinfo{author}{Kumar,
  P.}, \bibinfo{author}{Kim, K.}, \bibinfo{author}{Lee, S.},
  \bibinfo{author}{Bhattacharya, S.}, \bibinfo{year}{2019}.
\newblock \bibinfo{title}{Solid waste management: Scope and the challenge of
  sustainability}.
\newblock \bibinfo{journal}{Journal of cleaner production}
  \bibinfo{volume}{228}, \bibinfo{pages}{658--678}.
\newblock \DOIprefix\doi{10.1016/j.jclepro.2019.04.323}.
\bibitem[{De~Souza et~al.(2017)De~Souza, Gonz{\'a}lez, Faceli and
  Casadei}]{deX2017technologies}
\bibinfo{author}{De~Souza, A.}, \bibinfo{author}{Gonz{\'a}lez, S.},
  \bibinfo{author}{Faceli, K.}, \bibinfo{author}{Casadei, V.},
  \bibinfo{year}{2017}.
\newblock \bibinfo{title}{Technologies and decision support systems to aid
  solid-waste management: a systematic review}.
\newblock \bibinfo{journal}{Waste Management} \bibinfo{volume}{59},
  \bibinfo{pages}{567--584}.
\newblock \DOIprefix\doi{10.1016/j.wasman.2016.10.045}.
\bibitem[{Di~Felice(2014a)}]{di2014integration}
\bibinfo{author}{Di~Felice, P.}, \bibinfo{year}{2014}a.
\newblock \bibinfo{title}{Integration of spatial and descriptive information to
  solve the urban waste accumulation problem}.
\newblock \bibinfo{journal}{Procedia-Social and Behavioral Sciences}
  \bibinfo{volume}{147}, \bibinfo{pages}{182--188}.
\newblock \DOIprefix\doi{10.1016/j.sbspro.2014.07.150}.
\bibitem[{Di~Felice(2014b)}]{di2014integrationpilotcase}
\bibinfo{author}{Di~Felice, P.}, \bibinfo{year}{2014}b.
\newblock \bibinfo{title}{Integration of spatial and descriptive information to
  solve the urban waste accumulation problem: A pilot study}.
\newblock \bibinfo{journal}{Procedia-Social and Behavioral Sciences}
  \bibinfo{volume}{147}, \bibinfo{pages}{592--597}.
\newblock \DOIprefix\doi{10.1016/j.sbspro.2014.07.636}.
\bibitem[{Erfani et~al.(2017)Erfani, Danesh, Karrabi and
  Shad}]{erfani2017novel}
\bibinfo{author}{Erfani, S.}, \bibinfo{author}{Danesh, S.},
  \bibinfo{author}{Karrabi, S.}, \bibinfo{author}{Shad, R.},
  \bibinfo{year}{2017}.
\newblock \bibinfo{title}{A novel approach to find and optimize bin locations
  and collection routes using a geographic information system}.
\newblock \bibinfo{journal}{Waste Management \& Research} \bibinfo{volume}{35},
  \bibinfo{pages}{776--785}.
\newblock \DOIprefix\doi{10.1177/0734242X17706753}.
\bibitem[{Erfani et~al.(2018)Erfani, Danesh, Karrabi, Shad and
  Nemati}]{erfaniX2018using}
\bibinfo{author}{Erfani, S.}, \bibinfo{author}{Danesh, S.},
  \bibinfo{author}{Karrabi, S.}, \bibinfo{author}{Shad, R.},
  \bibinfo{author}{Nemati, S.}, \bibinfo{year}{2018}.
\newblock \bibinfo{title}{Using applied operations research and geographical
  information systems to evaluate effective factors in storage service of
  municipal solid waste management systems}.
\newblock \bibinfo{journal}{Waste Management} \bibinfo{volume}{79},
  \bibinfo{pages}{346--355}.
\newblock \DOIprefix\doi{10.1016/j.wasman.2018.08.003}.
\bibitem[{Eshet et~al.(2007)Eshet, Baron, Shechter and
  Ayalon}]{eshet2007measuring}
\bibinfo{author}{Eshet, T.}, \bibinfo{author}{Baron, M.},
  \bibinfo{author}{Shechter, M.}, \bibinfo{author}{Ayalon, O.},
  \bibinfo{year}{2007}.
\newblock \bibinfo{title}{Measuring externalities of waste transfer stations in
  israel using hedonic pricing}.
\newblock \bibinfo{journal}{Waste Management} \bibinfo{volume}{27},
  \bibinfo{pages}{614--625}.
\newblock \DOIprefix\doi{10.1016/j.wasman.2006.03.021}.
\bibitem[{Ferronato et~al.(2020)Ferronato, Preziosi, Portillo, Lizarazu and
  Torretta}]{ferronato2020assessment}
\bibinfo{author}{Ferronato, N.}, \bibinfo{author}{Preziosi, G.},
  \bibinfo{author}{Portillo, M.}, \bibinfo{author}{Lizarazu, E.},
  \bibinfo{author}{Torretta, V.}, \bibinfo{year}{2020}.
\newblock \bibinfo{title}{Assessment of municipal solid waste selective
  collection scenarios with geographic information systems in {Bolivia}}.
\newblock \bibinfo{journal}{Waste Management} \bibinfo{volume}{102},
  \bibinfo{pages}{919--931}.
\newblock \DOIprefix\doi{10.1016/j.wasman.2019.12.010}.
\bibitem[{Flahaut et~al.(2002)Flahaut, Laurent and
  Thomas}]{flahautX2002locating}
\bibinfo{author}{Flahaut, B.}, \bibinfo{author}{Laurent, M.},
  \bibinfo{author}{Thomas, I.}, \bibinfo{year}{2002}.
\newblock \bibinfo{title}{Locating a community recycling center within a
  residential area: a {B}elgian case study}.
\newblock \bibinfo{journal}{The Professional Geographer} \bibinfo{volume}{54},
  \bibinfo{pages}{67--82}.
\newblock \DOIprefix\doi{10.1111/0033-0124.00316}.
\bibitem[{Fleischmann et~al.(1997)Fleischmann, Bloemhof, Dekker, Van~der Laan,
  Van~Nunen and Van~Wassenhove}]{fleischmannX1997quantitative}
\bibinfo{author}{Fleischmann, M.}, \bibinfo{author}{Bloemhof, J.},
  \bibinfo{author}{Dekker, R.}, \bibinfo{author}{Van~der Laan, E.},
  \bibinfo{author}{Van~Nunen, J.}, \bibinfo{author}{Van~Wassenhove, L.},
  \bibinfo{year}{1997}.
\newblock \bibinfo{title}{Quantitative models for reverse logistics: A review}.
\newblock \bibinfo{journal}{European journal of operational research}
  \bibinfo{volume}{103}, \bibinfo{pages}{1--17}.
\newblock \DOIprefix\doi{10.1016/S0377-2217(97)00230-0}.
\bibitem[{Gallardo et~al.(2010)Gallardo, Bovea, Colomer, Prades and
  Carlos}]{gallardo2010comparison}
\bibinfo{author}{Gallardo, A.}, \bibinfo{author}{Bovea, M.},
  \bibinfo{author}{Colomer, F.}, \bibinfo{author}{Prades, M.},
  \bibinfo{author}{Carlos, M.}, \bibinfo{year}{2010}.
\newblock \bibinfo{title}{Comparison of different collection systems for sorted
  household waste in spain}.
\newblock \bibinfo{journal}{Waste Management} \bibinfo{volume}{30},
  \bibinfo{pages}{2430--2439}.
\newblock \DOIprefix\doi{10.1016/j.wasman.2010.05.026}.
\bibitem[{Gallardo et~al.(2015)Gallardo, Carlos, Peris and
  Colomer}]{gallardoX2015methodology}
\bibinfo{author}{Gallardo, A.}, \bibinfo{author}{Carlos, M.},
  \bibinfo{author}{Peris, M.}, \bibinfo{author}{Colomer, F.},
  \bibinfo{year}{2015}.
\newblock \bibinfo{title}{Methodology to design a municipal solid waste
  pre-collection system. {A} case study}.
\newblock \bibinfo{journal}{Waste Management} \bibinfo{volume}{36},
  \bibinfo{pages}{1--11}.
\newblock \DOIprefix\doi{10.1016/j.wasman.2014.11.008}.
\bibitem[{Gautam and Kumar(2005)}]{gautam2005strategic}
\bibinfo{author}{Gautam, A.}, \bibinfo{author}{Kumar, S.},
  \bibinfo{year}{2005}.
\newblock \bibinfo{title}{Strategic planning of recycling options by
  multi-objective programming in a {GIS} environment}.
\newblock \bibinfo{journal}{Clean Technologies and Environmental Policy}
  \bibinfo{volume}{7}, \bibinfo{pages}{306--316}.
\newblock \DOIprefix\doi{10.1007/s10098-005-0006-7}.
\bibitem[{Ghiani et~al.(2014a)Ghiani, Lagan{\`a}, Manni, Musmanno and
  Vigo}]{ghianiX2014operations}
\bibinfo{author}{Ghiani, G.}, \bibinfo{author}{Lagan{\`a}, D.},
  \bibinfo{author}{Manni, E.}, \bibinfo{author}{Musmanno, R.},
  \bibinfo{author}{Vigo, D.}, \bibinfo{year}{2014}a.
\newblock \bibinfo{title}{Operations research in solid waste management: A
  survey of strategic and tactical issues}.
\newblock \bibinfo{journal}{Computers \& Operations Research}
  \bibinfo{volume}{44}, \bibinfo{pages}{22--32}.
\newblock \DOIprefix\doi{10.1016/j.cor.2013.10.006}.
\bibitem[{Ghiani et~al.(2012)Ghiani, Lagan{\`a}, Manni and
  Triki}]{ghianiX2012capacitated}
\bibinfo{author}{Ghiani, G.}, \bibinfo{author}{Lagan{\`a}, D.},
  \bibinfo{author}{Manni, E.}, \bibinfo{author}{Triki, C.},
  \bibinfo{year}{2012}.
\newblock \bibinfo{title}{Capacitated location of collection sites in an urban
  waste management system}.
\newblock \bibinfo{journal}{Waste Management} \bibinfo{volume}{32},
  \bibinfo{pages}{1291--1296}.
\newblock \DOIprefix\doi{10.1016/j.wasman.2012.02.009}.
\bibitem[{Ghiani et~al.(2014b)Ghiani, Manni, Manni and
  Toraldo}]{ghianiX2014impact}
\bibinfo{author}{Ghiani, G.}, \bibinfo{author}{Manni, A.},
  \bibinfo{author}{Manni, E.}, \bibinfo{author}{Toraldo, M.},
  \bibinfo{year}{2014}b.
\newblock \bibinfo{title}{The impact of an efficient collection sites location
  on the zoning phase in municipal solid waste management}.
\newblock \bibinfo{journal}{Waste Management} \bibinfo{volume}{34},
  \bibinfo{pages}{1949--1956}.
\newblock \DOIprefix\doi{10.1016/j.wasman.2014.05.026}.
\bibitem[{Gilardino et~al.(2017)Gilardino, Rojas, Mattos, Larrea and
  V{\'a}zquez}]{gilardino2017combining}
\bibinfo{author}{Gilardino, A.}, \bibinfo{author}{Rojas, J.},
  \bibinfo{author}{Mattos, H.}, \bibinfo{author}{Larrea, G.},
  \bibinfo{author}{V{\'a}zquez, I.}, \bibinfo{year}{2017}.
\newblock \bibinfo{title}{Combining operational research and {Life Cycle
  Assessment} to optimize municipal solid waste collection in a district in
  {Lima (Peru)}}.
\newblock \bibinfo{journal}{Journal of Cleaner Production}
  \bibinfo{volume}{156}, \bibinfo{pages}{589--603}.
\newblock \DOIprefix\doi{10.1016/j.jclepro.2017.04.005}.
\bibitem[{Gonz{\'a}lez and Adenso(2002)}]{gonzalez2002model}
\bibinfo{author}{Gonz{\'a}lez, P.}, \bibinfo{author}{Adenso, B.},
  \bibinfo{year}{2002}.
\newblock \bibinfo{title}{A model for the reallocation of recycling containers:
  application to the case of glass}.
\newblock \bibinfo{journal}{Waste Management \& Research} \bibinfo{volume}{20},
  \bibinfo{pages}{398--406}.
\newblock \DOIprefix\doi{10.1177/0734242X0202000503}.
\bibitem[{Gough et~al.(2012)Gough, Oliver and Thomas}]{Gough2012}
\bibinfo{author}{Gough, D.}, \bibinfo{author}{Oliver, S.},
  \bibinfo{author}{Thomas, J.}, \bibinfo{year}{2012}.
\newblock \bibinfo{title}{An introduction to systematic reviews}.
\newblock \bibinfo{publisher}{SAGE}, \bibinfo{address}{London; Thousand Oaks,
  Calif.}
\bibitem[{Grant and Booth(2009)}]{Grant2009}
\bibinfo{author}{Grant, M.}, \bibinfo{author}{Booth, A.}, \bibinfo{year}{2009}.
\newblock \bibinfo{title}{A typology of reviews: an analysis of 14 review types
  and associated methodologies}.
\newblock \bibinfo{journal}{Health Information {\&} Libraries Journal}
  \bibinfo{volume}{26}, \bibinfo{pages}{91--108}.
\newblock \DOIprefix\doi{10.1111/j.1471-1842.2009.00848.x}.
\bibitem[{Gupta et~al.(2019)Gupta, Shree, Hiremath and
  Rajendran}]{gupta2019use}
\bibinfo{author}{Gupta, P.}, \bibinfo{author}{Shree, V.},
  \bibinfo{author}{Hiremath, L.}, \bibinfo{author}{Rajendran, S.},
  \bibinfo{year}{2019}.
\newblock \bibinfo{title}{The use of modern technology in smart waste
  management and recycling: Artificial intelligence and machine learning}, in:
  \bibinfo{editor}{Kumar, R.}, \bibinfo{editor}{Wiil, U.} (Eds.),
  \bibinfo{booktitle}{Recent Advances in Computational Intelligence}.
  \bibinfo{publisher}{Springer}, pp. \bibinfo{pages}{173--188}.
\newblock \DOIprefix\doi{10.1007/978-3-030-12500-4_11}.
\bibitem[{Hazra and Goel(2009)}]{hazra2009solid}
\bibinfo{author}{Hazra, T.}, \bibinfo{author}{Goel, S.}, \bibinfo{year}{2009}.
\newblock \bibinfo{title}{Solid waste management in {Kolkata, India}: Practices
  and challenges}.
\newblock \bibinfo{journal}{Waste Management} \bibinfo{volume}{29},
  \bibinfo{pages}{470--478}.
\newblock \DOIprefix\doi{10.1016/j.wasman.2008.01.023}.
\bibitem[{Hemmelmayr et~al.(2013)Hemmelmayr, Doerner, Hartl and
  Vigo}]{hemmelmayrX2013models}
\bibinfo{author}{Hemmelmayr, V.}, \bibinfo{author}{Doerner, K.},
  \bibinfo{author}{Hartl, R.}, \bibinfo{author}{Vigo, D.},
  \bibinfo{year}{2013}.
\newblock \bibinfo{title}{Models and algorithms for the integrated planning of
  bin allocation and vehicle routing in solid waste management}.
\newblock \bibinfo{journal}{Transportation Science} \bibinfo{volume}{48},
  \bibinfo{pages}{103--120}.
\newblock \DOIprefix\doi{10.1287/trsc.2013.0459}.
\bibitem[{Hemmelmayr et~al.(2017)Hemmelmayr, Smilowitz and De~la
  Torre}]{hemmelmayrX2017periodic}
\bibinfo{author}{Hemmelmayr, V.}, \bibinfo{author}{Smilowitz, K.},
  \bibinfo{author}{De~la Torre, L.}, \bibinfo{year}{2017}.
\newblock \bibinfo{title}{A periodic location routing problem for collaborative
  recycling}.
\newblock \bibinfo{journal}{IISE Transactions} \bibinfo{volume}{49},
  \bibinfo{pages}{414--428}.
\newblock \DOIprefix\doi{10.1080/24725854.2016.1267882}.
\bibitem[{Herrera et~al.(2018)Herrera, Imbaquingo, Lorente, Herrera, Peluffo
  and Rossit}]{herrera2018optimization}
\bibinfo{author}{Herrera, I.}, \bibinfo{author}{Imbaquingo, W.},
  \bibinfo{author}{Lorente, L.}, \bibinfo{author}{Herrera, E.},
  \bibinfo{author}{Peluffo, D.}, \bibinfo{author}{Rossit, D.},
  \bibinfo{year}{2018}.
\newblock \bibinfo{title}{Optimization of the network of urban solid waste
  containers: A case study}, in: \bibinfo{booktitle}{4th International
  Conference on Technology Trends}, pp. \bibinfo{pages}{578--589}.
\newblock \DOIprefix\doi{10.1007/978-3-030-05532-5_44}.
\bibitem[{Hoornweg and Bhada(2012)}]{hoornweg2012waste}
\bibinfo{author}{Hoornweg, D.}, \bibinfo{author}{Bhada, P.},
  \bibinfo{year}{2012}.
\newblock \bibinfo{title}{What a waste: a global review of solid waste
  management}.
\newblock \bibinfo{journal}{Urban development series} \bibinfo{volume}{15}.
\bibitem[{Hoornweg et~al.(2015)Hoornweg, Bhada and Kennedy}]{hoornweg2015peak}
\bibinfo{author}{Hoornweg, D.}, \bibinfo{author}{Bhada, P.},
  \bibinfo{author}{Kennedy, C.}, \bibinfo{year}{2015}.
\newblock \bibinfo{title}{Peak waste: When is it likely to occur?}
\newblock \bibinfo{journal}{Journal of Industrial Ecology}
  \bibinfo{volume}{19}, \bibinfo{pages}{117--128}.
\newblock \DOIprefix\doi{10.1111/jiec.12165}.
\bibitem[{Jammeli et~al.(2019)Jammeli, Argoubi and Masri}]{jammeliX2019bi}
\bibinfo{author}{Jammeli, H.}, \bibinfo{author}{Argoubi, M.},
  \bibinfo{author}{Masri, H.}, \bibinfo{year}{2019}.
\newblock \bibinfo{title}{A bi-objective stochastic programming model for the
  household waste collection and transportation problem: case of the city of
  {S}ousse}.
\newblock \bibinfo{journal}{Operational Research} ,
  \bibinfo{pages}{1--27}\DOIprefix\doi{10.1007/s12351-019-00538-5}.
  \bibinfo{note}{in press}.
\bibitem[{Kao and Lin(2002)}]{kaoX2002shortest}
\bibinfo{author}{Kao, J.}, \bibinfo{author}{Lin, T.}, \bibinfo{year}{2002}.
\newblock \bibinfo{title}{Shortest service location model for planning waste
  pickup locations}.
\newblock \bibinfo{journal}{Journal of the Air \& Waste Management Association}
  \bibinfo{volume}{52}, \bibinfo{pages}{585--592}.
\newblock \DOIprefix\doi{10.1080/10473289.2002.10470807}.
\bibitem[{Kao et~al.(2013)Kao, Tsai and Huang}]{kao2013spatial}
\bibinfo{author}{Kao, J.}, \bibinfo{author}{Tsai, Y.}, \bibinfo{author}{Huang,
  Y.}, \bibinfo{year}{2013}.
\newblock \bibinfo{title}{Spatial service location-allocation analysis for
  siting recycling depots}.
\newblock \bibinfo{journal}{Journal of Environmental Engineering}
  \bibinfo{volume}{139}, \bibinfo{pages}{1035--1041}.
\newblock \DOIprefix\doi{10.1061/(ASCE)EE.1943-7870.0000720}.
\bibitem[{Kao et~al.(2010)Kao, Wen and Liu}]{kao2010service}
\bibinfo{author}{Kao, J.}, \bibinfo{author}{Wen, L.}, \bibinfo{author}{Liu,
  K.}, \bibinfo{year}{2010}.
\newblock \bibinfo{title}{Service distance and ratio-based location-allocation
  models for siting recycling depots}.
\newblock \bibinfo{journal}{Journal of Environmental Engineering}
  \bibinfo{volume}{136}, \bibinfo{pages}{444--450}.
\newblock \DOIprefix\doi{10.1061/(ASCE)EE.1943-7870.0000170}.
\bibitem[{Karadimas et~al.(2005)Karadimas, Mavrantza and
  Loumos}]{karadimasX2005gis}
\bibinfo{author}{Karadimas, N.}, \bibinfo{author}{Mavrantza, O.},
  \bibinfo{author}{Loumos, V.}, \bibinfo{year}{2005}.
\newblock \bibinfo{title}{{GIS} integrated waste production modeling}, in:
  \bibinfo{booktitle}{EUROCON 2005 - The International Conference on ``Computer
  as a Tool''}, pp. \bibinfo{pages}{1279--1282}.
\newblock \DOIprefix\doi{10.1109/EURCON.2005.1630190}.
\bibitem[{Karadimas et~al.(2006)Karadimas, Orsoni and
  Loumos}]{karadimas2006municipal}
\bibinfo{author}{Karadimas, N.}, \bibinfo{author}{Orsoni, A.},
  \bibinfo{author}{Loumos, V.}, \bibinfo{year}{2006}.
\newblock \bibinfo{title}{Municipal solid waste generation modelling based on
  fuzzy logic}, in: \bibinfo{booktitle}{20th European Conference on Modelling
  and Simulation}, \bibinfo{address}{Bonn, Germany}.
\bibitem[{Karadimas and Loumos(2008)}]{karadimas2008gis}
\bibinfo{author}{Karadimas, N.V.}, \bibinfo{author}{Loumos, V.G.},
  \bibinfo{year}{2008}.
\newblock \bibinfo{title}{{GIS}-based modelling for the estimation of municipal
  solid waste generation and collection}.
\newblock \bibinfo{journal}{Waste Management \& Research} \bibinfo{volume}{26},
  \bibinfo{pages}{337--346}.
\newblock \DOIprefix\doi{10.1177/0734242X07081484}.
\bibitem[{Karkanias et~al.(2014)Karkanias, Perkoulidis, Grigoriadis, Stafylas,
  Dagdilelis, Feleki and Moussiopoulos}]{karkanias2014assessing}
\bibinfo{author}{Karkanias, C.}, \bibinfo{author}{Perkoulidis, G.},
  \bibinfo{author}{Grigoriadis, N.}, \bibinfo{author}{Stafylas, S.},
  \bibinfo{author}{Dagdilelis, E.}, \bibinfo{author}{Feleki, E.},
  \bibinfo{author}{Moussiopoulos, N.}, \bibinfo{year}{2014}.
\newblock \bibinfo{title}{Assessing recycling potential in local level: the
  case of {Neapoli-Sykies municipality, Greece}}, in:
  \bibinfo{booktitle}{Fresenius Environmental Bulletin}, pp.
  \bibinfo{pages}{2884--2889}.
\bibitem[{Khan and Samadder(2016)}]{khan2016allocation}
\bibinfo{author}{Khan, D.}, \bibinfo{author}{Samadder, S.},
  \bibinfo{year}{2016}.
\newblock \bibinfo{title}{{Allocation of solid waste collection bins and route
  optimisation using geographical information system: a case study of Dhanbad
  City, India}}.
\newblock \bibinfo{journal}{Waste Management \& Research} \bibinfo{volume}{34},
  \bibinfo{pages}{666--676}.
\newblock \DOIprefix\doi{10.1177/0734242X16649679}.
\bibitem[{Kim and Lee(2013)}]{kim2013restricted}
\bibinfo{author}{Kim, J.}, \bibinfo{author}{Lee, D.}, \bibinfo{year}{2013}.
\newblock \bibinfo{title}{A restricted dynamic model for refuse collection
  network design in reverse logistics}.
\newblock \bibinfo{journal}{Computers \& Industrial Engineering}
  \bibinfo{volume}{66}, \bibinfo{pages}{1131--1137}.
\newblock \DOIprefix\doi{10.1016/j.cie.2013.08.001}.
\bibitem[{Kim and Lee(2015a)}]{kim2015case}
\bibinfo{author}{Kim, J.}, \bibinfo{author}{Lee, D.}, \bibinfo{year}{2015}a.
\newblock \bibinfo{title}{A case study on collection network design, capacity
  planning and vehicle routing in reverse logistics}.
\newblock \bibinfo{journal}{International Journal of Sustainable Engineering}
  \bibinfo{volume}{8}, \bibinfo{pages}{66--76}.
\newblock \DOIprefix\doi{10.1080/19397038.2014.947393}.
\bibitem[{Kim and Lee(2015b)}]{kim2015integrated}
\bibinfo{author}{Kim, J.}, \bibinfo{author}{Lee, D.}, \bibinfo{year}{2015}b.
\newblock \bibinfo{title}{An integrated approach for collection network design,
  capacity planning and vehicle routing in reverse logistics}.
\newblock \bibinfo{journal}{Journal of the Operational Research Society}
  \bibinfo{volume}{66}, \bibinfo{pages}{76--85}.
\newblock \DOIprefix\doi{10.1057/jors.2013.168}.
\bibitem[{Kumar and Goel(2009)}]{kumar2009characterization}
\bibinfo{author}{Kumar, K.}, \bibinfo{author}{Goel, S.}, \bibinfo{year}{2009}.
\newblock \bibinfo{title}{{Characterization of municipal solid waste (MSW) and
  a proposed management plan for Kharagpur, West Bengal, India}}.
\newblock \bibinfo{journal}{Resources, Conservation and Recycling}
  \bibinfo{volume}{53}, \bibinfo{pages}{166--174}.
\newblock \DOIprefix\doi{10.1016/j.resconrec.2008.11.004}.
\bibitem[{Lebersorger and Beigl(2011)}]{lebersorger2011municipal}
\bibinfo{author}{Lebersorger, S.}, \bibinfo{author}{Beigl, P.},
  \bibinfo{year}{2011}.
\newblock \bibinfo{title}{Municipal solid waste generation in municipalities:
  Quantifying impacts of household structure, commercial waste and domestic
  fuel}.
\newblock \bibinfo{journal}{Waste management} \bibinfo{volume}{31},
  \bibinfo{pages}{1907--1915}.
\newblock \DOIprefix\doi{10.1016/j.wasman.2011.05.016}.
\bibitem[{Leeabai et~al.(2021)Leeabai, Areeprasert, Khaobang,
  Viriyapanitchakij, Bussa, Dilinazi and Takahashi}]{leeabai2021effects}
\bibinfo{author}{Leeabai, N.}, \bibinfo{author}{Areeprasert, C.},
  \bibinfo{author}{Khaobang, C.}, \bibinfo{author}{Viriyapanitchakij, N.},
  \bibinfo{author}{Bussa, B.}, \bibinfo{author}{Dilinazi, D.},
  \bibinfo{author}{Takahashi, F.}, \bibinfo{year}{2021}.
\newblock \bibinfo{title}{The effects of color preference and noticeability of
  trash bins on waste collection performance and waste-sorting behaviors}.
\newblock \bibinfo{journal}{Waste Management} \bibinfo{volume}{121},
  \bibinfo{pages}{153--163}.
\newblock \DOIprefix\doi{10.1016/j.wasman.2020.12.010}.
\bibitem[{Leeabai et~al.(2019)Leeabai, Suzuki, Jiang, Dilixiati and
  Takahashi}]{leeabai2019effects}
\bibinfo{author}{Leeabai, N.}, \bibinfo{author}{Suzuki, S.},
  \bibinfo{author}{Jiang, Q.}, \bibinfo{author}{Dilixiati, D.},
  \bibinfo{author}{Takahashi, F.}, \bibinfo{year}{2019}.
\newblock \bibinfo{title}{The effects of setting conditions of trash bins on
  waste collection performance and waste separation behaviors; distance from
  walking path, separated setting, and arrangements}.
\newblock \bibinfo{journal}{Waste Management} \bibinfo{volume}{94},
  \bibinfo{pages}{58--67}.
\newblock \DOIprefix\doi{10.1016/j.wasman.2019.05.039}.
\bibitem[{Letelier et~al.(2021)Letelier, Blazquez and
  Paredes}]{letelier2021solving}
\bibinfo{author}{Letelier, C.}, \bibinfo{author}{Blazquez, C.},
  \bibinfo{author}{Paredes, G.}, \bibinfo{year}{2021}.
\newblock \bibinfo{title}{{Solving the bin location--allocation problem for
  household and recycle waste generated in the commune of Renca in Santiago,
  Chile}}.
\newblock \bibinfo{journal}{Waste Management \& Research}
  \DOIprefix\doi{10.1177/0734242X20986610}. \bibinfo{note}{in press}.
\bibitem[{Li et~al.(2020)Li, Batty and Goodchild}]{li2020real}
\bibinfo{author}{Li, W.}, \bibinfo{author}{Batty, M.},
  \bibinfo{author}{Goodchild, M.}, \bibinfo{year}{2020}.
\newblock \bibinfo{title}{Real-time {GIS} for smart cities}.
\newblock \bibinfo{journal}{International Journal of Geographical Information
  Science} \bibinfo{volume}{34}, \bibinfo{pages}{311--324}.
\newblock \DOIprefix\doi{10.1080/13658816.2019.1673397}.
\bibitem[{Lin et~al.(2011)Lin, Tsai, Chen and Kao}]{linX2011model}
\bibinfo{author}{Lin, H.}, \bibinfo{author}{Tsai, Z.}, \bibinfo{author}{Chen,
  G.}, \bibinfo{author}{Kao, J.}, \bibinfo{year}{2011}.
\newblock \bibinfo{title}{A model for the implementation of a two-shift
  municipal solid waste and recyclable material collection plan that offers
  greater convenience to residents}.
\newblock \bibinfo{journal}{Journal of the Air \& Waste Management Association}
  \bibinfo{volume}{61}, \bibinfo{pages}{55--62}.
\newblock \DOIprefix\doi{10.3155/1047-3289.61.1.55}.
\bibitem[{Lindell and Earle(1983)}]{lindell1983close}
\bibinfo{author}{Lindell, M.}, \bibinfo{author}{Earle, T.},
  \bibinfo{year}{1983}.
\newblock \bibinfo{title}{How close is close enough: Public perceptions of the
  risks of industrial facilities}.
\newblock \bibinfo{journal}{Risk Analysis} \bibinfo{volume}{3},
  \bibinfo{pages}{245--253}.
\newblock \DOIprefix\doi{10.1111/j.1539-6924.1983.tb01393.x}.
\bibitem[{L{\'o}pez et~al.(2008)L{\'o}pez, Aguilar, Fern{\'a}ndez and
  Jim{\'e}nez~del Valle}]{lopez2008optimizing}
\bibinfo{author}{L{\'o}pez, J.}, \bibinfo{author}{Aguilar, M.},
  \bibinfo{author}{Fern{\'a}ndez, S.}, \bibinfo{author}{Jim{\'e}nez~del Valle,
  A.}, \bibinfo{year}{2008}.
\newblock \bibinfo{title}{Optimizing the collection of used paper from small
  businesses through gis techniques: The legan{\'e}s case (madrid, spain)}.
\newblock \bibinfo{journal}{Waste Management} \bibinfo{volume}{28},
  \bibinfo{pages}{282--293}.
\newblock \DOIprefix\doi{10.1016/j.wasman.2007.02.036}.
\bibitem[{L{\'o}pez et~al.(2009)L{\'o}pez, Aguilar, Soriano and Fernando~de
  Fuentes}]{lopez2009containerisation}
\bibinfo{author}{L{\'o}pez, J.}, \bibinfo{author}{Aguilar, M.},
  \bibinfo{author}{Soriano, F.}, \bibinfo{author}{Fernando~de Fuentes, A.},
  \bibinfo{year}{2009}.
\newblock \bibinfo{title}{Containerisation of the selective collection of light
  packaging waste material: The case of small cities in advanced economies}.
\newblock \bibinfo{journal}{Cities} \bibinfo{volume}{26},
  \bibinfo{pages}{339--348}.
\newblock \DOIprefix\doi{10.1016/j.cities.2009.09.002}.
\bibitem[{Lu and Nimbole(2008)}]{lu2008intro}
\bibinfo{author}{Lu, H.}, \bibinfo{author}{Nimbole, P.}, \bibinfo{year}{2008}.
\newblock \bibinfo{title}{Intro to TransCAD GIS}.
\newblock \bibinfo{type}{Technical Report}. Connect NCDOT Business Partner
  Resources.
\bibitem[{Maraqa et~al.(2018)Maraqa, Aldahab, Ghanma and
  Al~Kaabi}]{maraqa2018optimization}
\bibinfo{author}{Maraqa, M.}, \bibinfo{author}{Aldahab, E.},
  \bibinfo{author}{Ghanma, M.}, \bibinfo{author}{Al~Kaabi, S.},
  \bibinfo{year}{2018}.
\newblock \bibinfo{title}{Optimization of fuel consumption for municipal solid
  waste collection in {Al Ain city, UAE}}, in: \bibinfo{booktitle}{2018
  International Joint Conference on Materials Science and Mechanical
  Engineering}, pp. \bibinfo{pages}{12--26}.
\newblock \DOIprefix\doi{10.1088/1757-899X/383/1/012026}.
\bibitem[{Nesmachnow(2014)}]{Nesmachnow2014}
\bibinfo{author}{Nesmachnow, S.}, \bibinfo{year}{2014}.
\newblock \bibinfo{title}{An overview of metaheuristics: accurate and efficient
  methods for optimisation}.
\newblock \bibinfo{journal}{International Journal of Metaheuristics}
  \bibinfo{volume}{3}, \bibinfo{pages}{320--347}.
\bibitem[{Nesmachnow et~al.(2018)Nesmachnow, Rossit and
  Toutouh}]{nesmachnow2018comparison}
\bibinfo{author}{Nesmachnow, S.}, \bibinfo{author}{Rossit, D.},
  \bibinfo{author}{Toutouh, J.}, \bibinfo{year}{2018}.
\newblock \bibinfo{title}{Comparison of multiobjective evolutionary algorithms
  for prioritized urban waste collection in montevideo, uruguay}.
\newblock \bibinfo{journal}{Electronic Notes in Discrete Mathematics}
  \bibinfo{volume}{69}, \bibinfo{pages}{93--100}.
\newblock \DOIprefix\doi{10.1016/j.endm.2018.07.013}.
\bibitem[{Nevrl{\`y} et~al.(2019)Nevrl{\`y}, {\v{S}}ompl{\'a}k, Khyr,
  Smejkalova and Jadrn{\`y}}]{nevrly2019municipal}
\bibinfo{author}{Nevrl{\`y}, V.}, \bibinfo{author}{{\v{S}}ompl{\'a}k, R.},
  \bibinfo{author}{Khyr, L.}, \bibinfo{author}{Smejkalova, V.},
  \bibinfo{author}{Jadrn{\`y}, J.}, \bibinfo{year}{2019}.
\newblock \bibinfo{title}{Municipal solid waste container location based on
  walking distance and distribution of population}.
\newblock \bibinfo{journal}{Chemical Engineering Transactions}
  \bibinfo{volume}{76}, \bibinfo{pages}{553--558}.
\newblock \DOIprefix\doi{10.3303/CET1976093}.
\bibitem[{Nevrl{\`y} et~al.(2021)Nevrl{\`y}, {\v{S}}ompl{\'a}k, Smejkalova,
  Lipovsk{\`y} and Jadrn{\`y}}]{nevrly2021location}
\bibinfo{author}{Nevrl{\`y}, V.}, \bibinfo{author}{{\v{S}}ompl{\'a}k, R.},
  \bibinfo{author}{Smejkalova, V.}, \bibinfo{author}{Lipovsk{\`y}, T.},
  \bibinfo{author}{Jadrn{\`y}, J.}, \bibinfo{year}{2021}.
\newblock \bibinfo{title}{Location of municipal waste containers: Trade-off
  between criteria}.
\newblock \bibinfo{journal}{Journal of Cleaner Production}
  \bibinfo{volume}{278}, \bibinfo{pages}{123445}.
\newblock \DOIprefix\doi{10.1016/j.jclepro.2020.123445}.
\bibitem[{Nithya et~al.(2012)Nithya, Velumani and Kumar}]{nithya2012optimal}
\bibinfo{author}{Nithya, R.}, \bibinfo{author}{Velumani, A.},
  \bibinfo{author}{Kumar, S.}, \bibinfo{year}{2012}.
\newblock \bibinfo{title}{Optimal location and proximity distance of municipal
  solid waste collection bin using {GIS}: A case study of {Coimbatore} city}.
\newblock \bibinfo{journal}{WSEAS Transactions on environment and development}
  \bibinfo{volume}{8}, \bibinfo{pages}{107--119}.
\bibitem[{Oliaei and Fataei(2016)}]{oliaeibreakdown}
\bibinfo{author}{Oliaei, A.}, \bibinfo{author}{Fataei, E.},
  \bibinfo{year}{2016}.
\newblock \bibinfo{title}{Breakdown of urban waste repository location using
  {GIS (Case study District 3 region 1 Tabriz)}}.
\newblock \bibinfo{journal}{Ecology, Environment and Conservation}
  \bibinfo{volume}{22}, \bibinfo{pages}{551--557}.
\bibitem[{Parrot et~al.(2009)Parrot, Sotamenou and Dia}]{parrotX2009municipal}
\bibinfo{author}{Parrot, L.}, \bibinfo{author}{Sotamenou, J.},
  \bibinfo{author}{Dia, B.}, \bibinfo{year}{2009}.
\newblock \bibinfo{title}{{Municipal solid waste management in Africa:
  Strategies and livelihoods in Yaound{\'e}, Cameroon}}.
\newblock \bibinfo{journal}{Waste Management} \bibinfo{volume}{29},
  \bibinfo{pages}{986--995}.
\newblock \DOIprefix\doi{10.1016/j.wasman.2008.05.005}.
\bibitem[{Paul et~al.(2017)Paul, Dutta and Krishna}]{paul2017using}
\bibinfo{author}{Paul, K.}, \bibinfo{author}{Dutta, A.},
  \bibinfo{author}{Krishna, A.}, \bibinfo{year}{2017}.
\newblock \bibinfo{title}{Using {GIS} to locate waste bins: a case study on
  {Kolkata City, India}}.
\newblock \bibinfo{journal}{{Journal of Environmental Science and Management}}
  \bibinfo{volume}{20}.
\bibitem[{P{\'e}rez et~al.(2017)P{\'e}rez, Lumbreras, De~la Paz and
  Rodr{\'\i}guez}]{perez2017methodology}
\bibinfo{author}{P{\'e}rez, J.}, \bibinfo{author}{Lumbreras, J.},
  \bibinfo{author}{De~la Paz, D.}, \bibinfo{author}{Rodr{\'\i}guez, E.},
  \bibinfo{year}{2017}.
\newblock \bibinfo{title}{Methodology to evaluate the environmental impact of
  urban solid waste containerization system: A case study}.
\newblock \bibinfo{journal}{Journal of Cleaner Production}
  \bibinfo{volume}{150}, \bibinfo{pages}{197--213}.
\newblock \DOIprefix\doi{10.1016/j.jclepro.2017.03.003}.
\bibitem[{Purkayastha et~al.(2015)Purkayastha, Majumder and
  Chakrabarti}]{purkayastha2015collection}
\bibinfo{author}{Purkayastha, D.}, \bibinfo{author}{Majumder, M.},
  \bibinfo{author}{Chakrabarti, S.}, \bibinfo{year}{2015}.
\newblock \bibinfo{title}{Collection and recycle bin location-allocation
  problem in solid waste management: A review}.
\newblock \bibinfo{journal}{Pollution} \bibinfo{volume}{1},
  \bibinfo{pages}{175--191}.
\bibitem[{Rathore and Sarmah(2019)}]{rathore2019allocation}
\bibinfo{author}{Rathore, P.}, \bibinfo{author}{Sarmah, S.},
  \bibinfo{year}{2019}.
\newblock \bibinfo{title}{Allocation of bins in urban solid waste logistics
  system}, in: \bibinfo{booktitle}{4th International Conference on Harmony
  Search, Soft Computing and Applications}, pp. \bibinfo{pages}{485--495}.
\newblock \DOIprefix\doi{10.1007/978-981-13-0761-4_47}.
\bibitem[{Rathore et~al.(2020)Rathore, Sarmah and Singh}]{rathore2020location}
\bibinfo{author}{Rathore, P.}, \bibinfo{author}{Sarmah, S.},
  \bibinfo{author}{Singh, A.}, \bibinfo{year}{2020}.
\newblock \bibinfo{title}{Location--allocation of bins in urban solid waste
  management: a case study of {Bilaspur city, India}}.
\newblock \bibinfo{journal}{Environment, Development and Sustainability}
  \bibinfo{volume}{22}, \bibinfo{pages}{3309--3331}.
\newblock \DOIprefix\doi{10.1007/s10668-019-00347-y}.
\bibitem[{Ratkovi{\'c} et~al.(2016)Ratkovi{\'c}, Popovi{\'c}, Radivojevi{\'c}
  and Bjeli{\'c}}]{ratkovic2016planning}
\bibinfo{author}{Ratkovi{\'c}, B.}, \bibinfo{author}{Popovi{\'c}, D.},
  \bibinfo{author}{Radivojevi{\'c}, G.}, \bibinfo{author}{Bjeli{\'c}, N.},
  \bibinfo{year}{2016}.
\newblock \bibinfo{title}{Planning logistics network for recyclables
  collection}.
\newblock \bibinfo{journal}{Yugoslav Journal of Operations Research}
  \bibinfo{volume}{24}, \bibinfo{pages}{371--381}.
\newblock \DOIprefix\doi{10.2298/YJOR140408022R}.
\bibitem[{Rossit(2018)}]{rossit2018thesis}
\bibinfo{author}{Rossit, D.}, \bibinfo{year}{2018}.
\newblock \bibinfo{title}{Desarrollo de modelos y algoritmos para optimizar
  redes log\'{i}sticas de residuos s\'{o}lidos urbanos}.
\newblock Ph.D. thesis. Department of Engineering, Universidad Nacional del
  Sur. \bibinfo{address}{Bah\'{i}a Blanca, Argentina}.
\bibitem[{Rossit et~al.(2018)Rossit, Nesmachnow and
  Toutouh}]{rossit2018municipal}
\bibinfo{author}{Rossit, D.}, \bibinfo{author}{Nesmachnow, S.},
  \bibinfo{author}{Toutouh, J.}, \bibinfo{year}{2018}.
\newblock \bibinfo{title}{Municipal solid waste management in smart cities:
  facility location of community bins}, in: \bibinfo{booktitle}{First
  Ibero-American Congress on Information Management and Big Data}, pp.
  \bibinfo{pages}{102--115}.
\newblock \DOIprefix\doi{10.1007/978-3-030-12804-3_9}.
\bibitem[{Rossit et~al.(2019)Rossit, Nesmachnow and Toutouh}]{rossit2019bi}
\bibinfo{author}{Rossit, D.}, \bibinfo{author}{Nesmachnow, S.},
  \bibinfo{author}{Toutouh, J.}, \bibinfo{year}{2019}.
\newblock \bibinfo{title}{A bi-objective integer programming model for locating
  garbage accumulation points: a case study}.
\newblock \bibinfo{journal}{Revista Facultad de Ingenier{\'\i}a Universidad de
  Antioquia} ,
  \bibinfo{pages}{70--81}\DOIprefix\doi{10.17533/udea.redin.20190509}.
\bibitem[{Rossit et~al.(2017)Rossit, Tohm{\'e}, Frutos and
  Broz}]{rossitX2017application}
\bibinfo{author}{Rossit, D.}, \bibinfo{author}{Tohm{\'e}, F.},
  \bibinfo{author}{Frutos, M.}, \bibinfo{author}{Broz, D.},
  \bibinfo{year}{2017}.
\newblock \bibinfo{title}{An application of the augmented
  $\varepsilon$-constraint method to design a municipal sorted waste collection
  system}.
\newblock \bibinfo{journal}{Decision Science Letters} \bibinfo{volume}{6},
  \bibinfo{pages}{323--336}.
\newblock \DOIprefix\doi{10.5267/j.dsl.2017.3.001}.
\bibitem[{Rossit et~al.(2020)Rossit, Toutouh and Nesmachnow}]{rossit2020exact}
\bibinfo{author}{Rossit, D.}, \bibinfo{author}{Toutouh, J.},
  \bibinfo{author}{Nesmachnow, S.}, \bibinfo{year}{2020}.
\newblock \bibinfo{title}{Exact and heuristic approaches for multi-objective
  garbage accumulation points location in real scenarios}.
\newblock \bibinfo{journal}{Waste Management} \bibinfo{volume}{105},
  \bibinfo{pages}{467--481}.
\newblock \DOIprefix\doi{10.1016/j.wasman.2020.02.016}.
\bibitem[{Sheriff et~al.(2017)Sheriff, Subramanian, Rahman and
  Jayaram}]{sheriff2017integrated}
\bibinfo{author}{Sheriff, K.}, \bibinfo{author}{Subramanian, N.},
  \bibinfo{author}{Rahman, S.}, \bibinfo{author}{Jayaram, J.},
  \bibinfo{year}{2017}.
\newblock \bibinfo{title}{Integrated optimization model and methodology for
  plastics recycling: {I}ndian empirical evidence}.
\newblock \bibinfo{journal}{Journal of Cleaner Production}
  \bibinfo{volume}{153}, \bibinfo{pages}{707--717}.
\newblock \DOIprefix\doi{10.1016/j.jclepro.2016.07.137}.
\bibitem[{Singh(2019)}]{singh2019managing}
\bibinfo{author}{Singh, A.}, \bibinfo{year}{2019}.
\newblock \bibinfo{title}{Managing the uncertainty problems of municipal solid
  waste disposal}.
\newblock \bibinfo{journal}{Journal of Environmental Management}
  \bibinfo{volume}{240}, \bibinfo{pages}{259--265}.
\newblock \DOIprefix\doi{10.1016/j.compchemeng.2018.12.022}.
\bibitem[{Snyder(2019)}]{Snyder2019}
\bibinfo{author}{Snyder, H.}, \bibinfo{year}{2019}.
\newblock \bibinfo{title}{Literature review as a research methodology: An
  overview and guidelines}.
\newblock \bibinfo{journal}{Journal of Business Research}
  \bibinfo{volume}{104}, \bibinfo{pages}{333--339}.
\newblock \DOIprefix\doi{10.1016/j.jbusres.2019.07.039}.
\bibitem[{Sotamenou et~al.(2019)Sotamenou, De~Jaeger and
  Rousseau}]{sotamenou2019drivers}
\bibinfo{author}{Sotamenou, J.}, \bibinfo{author}{De~Jaeger, S.},
  \bibinfo{author}{Rousseau, S.}, \bibinfo{year}{2019}.
\newblock \bibinfo{title}{Drivers of legal and illegal solid waste disposal in
  the global south-the case of households in {Y}aound{\'e} ({C}ameroon)}.
\newblock \bibinfo{journal}{Journal of Environmental Management}
  \bibinfo{volume}{240}, \bibinfo{pages}{321--330}.
\newblock \DOIprefix\doi{10.1016/j.jenvman.2019.03.098}.
\bibitem[{Toutouh et~al.(2018)Toutouh, Rossit and
  Nesmachnow}]{toutouh2018intelligence}
\bibinfo{author}{Toutouh, J.}, \bibinfo{author}{Rossit, D.},
  \bibinfo{author}{Nesmachnow, S.}, \bibinfo{year}{2018}.
\newblock \bibinfo{title}{Computational intelligence for locating garbage
  accumulation points in urban scenarios}, in: \bibinfo{booktitle}{12th
  International Conference on Learning and Intelligent Optimization}, pp.
  \bibinfo{pages}{411--426}.
\newblock \DOIprefix\doi{10.1007/978-3-030-05348-2_34}.
\bibitem[{Toutouh et~al.(2020)Toutouh, Rossit and Nesmachnow}]{toutouh2020soft}
\bibinfo{author}{Toutouh, J.}, \bibinfo{author}{Rossit, D.},
  \bibinfo{author}{Nesmachnow, S.}, \bibinfo{year}{2020}.
\newblock \bibinfo{title}{Soft computing methods for multiobjective location of
  garbage accumulation points in smart cities}.
\newblock \bibinfo{journal}{Annals of Mathematics and Artificial Intelligence}
  \bibinfo{volume}{88}, \bibinfo{pages}{105--131}.
\newblock \DOIprefix\doi{10.1007/s10472-019-09647-5}.
\bibitem[{Tralh{\~a}o et~al.(2010)Tralh{\~a}o, Coutinho and
  Al{\c{c}}ada}]{tralhaoX2010multiobjective}
\bibinfo{author}{Tralh{\~a}o, L.}, \bibinfo{author}{Coutinho, J.},
  \bibinfo{author}{Al{\c{c}}ada, L.}, \bibinfo{year}{2010}.
\newblock \bibinfo{title}{A multiobjective modeling approach to locate
  multi-compartment containers for urban-sorted waste}.
\newblock \bibinfo{journal}{Waste Management} \bibinfo{volume}{30},
  \bibinfo{pages}{2418--2429}.
\newblock \DOIprefix\doi{10.1016/j.wasman.2010.06.017}.
\bibitem[{Tranfield et~al.(2003)Tranfield, Denyer and Smart}]{Tranfield2003}
\bibinfo{author}{Tranfield, D.}, \bibinfo{author}{Denyer, D.},
  \bibinfo{author}{Smart, P.}, \bibinfo{year}{2003}.
\newblock \bibinfo{title}{Towards a methodology for developing
  evidence-informed management knowledge by means of systematic review}.
\newblock \bibinfo{journal}{British Journal of Management}
  \bibinfo{volume}{14}, \bibinfo{pages}{207--222}.
\newblock \DOIprefix\doi{10.1111/1467-8551.00375}.
\bibitem[{Ugwuishiwu et~al.(2020)Ugwuishiwu, Nwoke, Okechukwu and
  Echiegu}]{ugwuishiwu2020gis}
\bibinfo{author}{Ugwuishiwu, B.}, \bibinfo{author}{Nwoke, O.},
  \bibinfo{author}{Okechukwu, C.}, \bibinfo{author}{Echiegu, E.},
  \bibinfo{year}{2020}.
\newblock \bibinfo{title}{{GIS}-based system analysis for waste bin location in
  {E}nugu municipality}.
\newblock \bibinfo{journal}{Agricultural Engineering International: CIGR
  Journal} \bibinfo{volume}{22}, \bibinfo{pages}{250--259}.
\bibitem[{Valeo et~al.(1998)Valeo, Baetz and Tsanis}]{valeo1998location}
\bibinfo{author}{Valeo, C.}, \bibinfo{author}{Baetz, B.W.},
  \bibinfo{author}{Tsanis, I.}, \bibinfo{year}{1998}.
\newblock \bibinfo{title}{Location of recycling depots with {GIS}}.
\newblock \bibinfo{journal}{Journal of Urban Planning and Development}
  \bibinfo{volume}{124}, \bibinfo{pages}{93--99}.
\newblock \DOIprefix\doi{10.1061/(ASCE)0733-9488(1998)124:2(93)}.
\bibitem[{Van~Engeland et~al.(2020)Van~Engeland, Beli{\"e}n, De~Boeck and
  De~Jaeger}]{van2020literature}
\bibinfo{author}{Van~Engeland, J.}, \bibinfo{author}{Beli{\"e}n, J.},
  \bibinfo{author}{De~Boeck, L.}, \bibinfo{author}{De~Jaeger, S.},
  \bibinfo{year}{2020}.
\newblock \bibinfo{title}{Literature review: Strategic network optimization
  models in waste reverse supply chains}.
\newblock \bibinfo{journal}{Omega} \bibinfo{volume}{91},
  \bibinfo{pages}{102012}.
\newblock \DOIprefix\doi{10.1016/j.omega.2018.12.001}.
\bibitem[{Vidovi{\'c} et~al.(2016)Vidovi{\'c}, Ratkovi{\'c}, Bjeli{\'c} and
  Popovi{\'c}}]{vidovic2016two}
\bibinfo{author}{Vidovi{\'c}, M.}, \bibinfo{author}{Ratkovi{\'c}, B.},
  \bibinfo{author}{Bjeli{\'c}, N.}, \bibinfo{author}{Popovi{\'c}, D.},
  \bibinfo{year}{2016}.
\newblock \bibinfo{title}{A two-echelon location-routing model for designing
  recycling logistics networks with profit: {MILP} and heuristic approach}.
\newblock \bibinfo{journal}{Expert Systems with Applications}
  \bibinfo{volume}{51}, \bibinfo{pages}{34--48}.
\newblock \DOIprefix\doi{10.1016/j.eswa.2015.12.029}.
\bibitem[{Vijay et~al.(2008)Vijay, Gautam, Kalamdhad, Gupta and
  Devotta}]{vijay2008gis}
\bibinfo{author}{Vijay, R.}, \bibinfo{author}{Gautam, A.},
  \bibinfo{author}{Kalamdhad, A.}, \bibinfo{author}{Gupta, A.},
  \bibinfo{author}{Devotta, S.}, \bibinfo{year}{2008}.
\newblock \bibinfo{title}{{GIS}-based locational analysis of collection bins in
  municipal solid waste management systems}.
\newblock \bibinfo{journal}{Journal of Environmental Engineering and Science}
  \bibinfo{volume}{7}, \bibinfo{pages}{39--43}.
\newblock \DOIprefix\doi{10.1139/S07-033}.
\bibitem[{Vijay et~al.(2005)Vijay, Gupta, Kalamdhad and
  Devotta}]{vijay2005estimation}
\bibinfo{author}{Vijay, R.}, \bibinfo{author}{Gupta, A.},
  \bibinfo{author}{Kalamdhad, A.}, \bibinfo{author}{Devotta, S.},
  \bibinfo{year}{2005}.
\newblock \bibinfo{title}{Estimation and allocation of solid waste to bin
  through geographical information systems}.
\newblock \bibinfo{journal}{Waste Management \& Research} \bibinfo{volume}{23},
  \bibinfo{pages}{479--484}.
\newblock \DOIprefix\doi{10.1177/0734242X05057763}.
\bibitem[{Vu et~al.(2018)Vu, Ng and Bolingbroke}]{vu2018parameter}
\bibinfo{author}{Vu, H.}, \bibinfo{author}{Ng, K.},
  \bibinfo{author}{Bolingbroke, D.}, \bibinfo{year}{2018}.
\newblock \bibinfo{title}{Parameter interrelationships in a dual phase
  {GIS}-based municipal solid waste collection model}.
\newblock \bibinfo{journal}{Waste Management} \bibinfo{volume}{78},
  \bibinfo{pages}{258--270}.
\newblock \DOIprefix\doi{10.1016/j.wasman.2018.05.050}.
\bibitem[{Yaakoubi et~al.(2018)Yaakoubi, Benabdouallah and
  Bojji}]{yaakoubi2018heuristic}
\bibinfo{author}{Yaakoubi, O.}, \bibinfo{author}{Benabdouallah, M.},
  \bibinfo{author}{Bojji, C.}, \bibinfo{year}{2018}.
\newblock \bibinfo{title}{Heuristic approaches for waste containers location
  problem and waste collection routes optimisation in an urban area}.
\newblock \bibinfo{journal}{International Journal of Environment and Waste
  Management} \bibinfo{volume}{21}, \bibinfo{pages}{269--286}.
\newblock \DOIprefix\doi{10.1504/IJEWM.2018.093436}.
\bibitem[{Yadav et~al.(2016)Yadav, Karmakar, Dikshit and
  Vanjari}]{yadav2016feasibility}
\bibinfo{author}{Yadav, V.}, \bibinfo{author}{Karmakar, S.},
  \bibinfo{author}{Dikshit, A.}, \bibinfo{author}{Vanjari, S.},
  \bibinfo{year}{2016}.
\newblock \bibinfo{title}{A feasibility study for the locations of waste
  transfer stations in urban centers: a case study on the city of nashik,
  india}.
\newblock \bibinfo{journal}{Journal of cleaner production}
  \bibinfo{volume}{126}, \bibinfo{pages}{191--205}.
\newblock \DOIprefix\doi{10.1016/j.jclepro.2016.03.017}.
\bibitem[{Zahan and Hasan(2020)}]{zahan2020multi}
\bibinfo{author}{Zahan, N.}, \bibinfo{author}{Hasan, Z.}, \bibinfo{year}{2020}.
\newblock \bibinfo{title}{Multi-attribute decision making approach for waste
  bin site selection problem: A case study in {Dhaka} city}, in:
  \bibinfo{booktitle}{11th International Conference on Computing, Communication
  and Networking Technologies}.
\newblock \DOIprefix\doi{10.1109/ICCCNT49239.2020.9225597}.
\bibitem[{Zamorano et~al.(2009)Zamorano, Molero, Grindlay, Rodr{\'\i}guez,
  Hurtado and Calvo}]{zamorano2009planning}
\bibinfo{author}{Zamorano, M.}, \bibinfo{author}{Molero, E.},
  \bibinfo{author}{Grindlay, A.}, \bibinfo{author}{Rodr{\'\i}guez, M.},
  \bibinfo{author}{Hurtado, A.}, \bibinfo{author}{Calvo, F.},
  \bibinfo{year}{2009}.
\newblock \bibinfo{title}{A planning scenario for the application of
  geographical information systems in municipal waste collection: {A case of
  Churriana de la Vega (Granada, Spain)}}.
\newblock \bibinfo{journal}{Resources, Conservation and Recycling}
  \bibinfo{volume}{54}, \bibinfo{pages}{123--133}.
\newblock \DOIprefix\doi{10.1016/j.resconrec.2009.07.001}.
\bibitem[{Zhang et~al.(2021)Zhang, Qin, Li and Tseng}]{zhang2021sustainable}
\bibinfo{author}{Zhang, J.}, \bibinfo{author}{Qin, Q.}, \bibinfo{author}{Li,
  G.}, \bibinfo{author}{Tseng, C.}, \bibinfo{year}{2021}.
\newblock \bibinfo{title}{Sustainable municipal waste management strategies
  through life cycle assessment method: A review}.
\newblock \bibinfo{journal}{Journal of Environmental Management}
  \bibinfo{volume}{287}, \bibinfo{pages}{112238}.
\newblock \DOIprefix\doi{10.1016/j.jenvman.2021.112238}.

\end{thebibliography}

\end{document}